\documentclass[twoside,12pt]{article}
\usepackage{epsfig}
\usepackage{epstopdf}
\newcommand{\be}{\begin{equation}}
\newcommand{\ee}{\end{equation}}
\newcommand{\bea}{\begin{eqnarray}}
\newcommand{\eea}{\end{eqnarray}}

\begin{document}

\title{ \vspace{1cm} Color Transparency: past, present and future}
\author{D.\ Dutta,$^{1}$ K.\ Hafidi,$^2$ M.\ Strikman$^3$\\
$^1$Mississippi State University, Mississippi State, MS 39762, USA\\
$^2$Argonne National Laboratory, Argonne, IL 60439, USA\\
$^3$Pennsylvania State University, University Park, PA 16802, USA\\ }
\maketitle
\begin{abstract}
We review a unique prediction of Quantum Chromo Dynamics, called color transparency (CT), where the final (and/or initial) state interactions of hadrons with the nuclear medium must vanish for exclusive processes at high momentum transfers. We retrace the progress of our understanding of this phenomenon, which began with the discovery of the $J/\psi$ meson,  followed by the discovery  of high energy CT phenomena, the recent developments in the investigations of the onset of CT at intermediate energies and the directions for future studies. 
\end{abstract}
\tableofcontents

\section{Introduction}
One of the distinctive properties of Quantum Chromo Dynamics (QCD) is the suppression of the interaction between small size color singlet wave packets and hadrons. It plays a key role in ensuring approximate Bjorken scaling in deep inelastic scattering, in proving QCD factorization theorems for high energy hard exclusive processes, and in the search for QCD related phenomena in nuclei. Moreover, it leads to a number of phenomena in the hard coherent/quasi-elastic interactions with nuclei at high energies that are dubbed as color transparency (CT). The phenomenon of CT refers to the cancellation of the QCD color fields for physically small singlet systems of quarks and gluons, which leads to the vanishing of the final (and/or initial) state interactions of hadrons with the  nuclear medium in exclusive processes at high momentum transfers. The simplest  model in which this phenomenon is present is the Low-Nussinov two-gluon exchange model~\cite{Low:1975sv,Nussinov:1975mw}. At intermediate energies CT phenomenon provides a unique probe of the space-time evolution of wave packets, which are relevant for the interpretation of the data from relativistic heavy ion collisions. For example interpretation of the baryon to pion ratio as a function of the transverse momentum, $p_t$, depends on whether a baryon can traverse a large 
distance while in a small size configuration as suggested in Ref.~\cite{Brodsky:2008qp}, and fluctuations of the nucleon to small and large size configurations lead to significant deviations from the Glauber picture in proton (nucleus) - nucleus collisions~\cite{Baym:1995cz}. In addition, the CT phenomenon is also a probe of the minimal small size components in the hadron's wave function. We review the phenomenon of CT, the conditions necessary for it to occur, theoretical and experimental progress made over the last few decades and the future outlook.

The story of CT goes back to the discovery of the $J/\psi$ meson. It was impossible to explain within the concepts of the pre-QCD theory of strong interactions why the decay width of $J/\psi$ is very narrow, and why the photoproduction cross section is so small. It was argued as early as the fall of 1974 \cite{Frankfurt:1975sy} that the radius of a system consisting of heavy quarks should be significantly smaller than the one given by the radius of pion emission. This was in contrast to the widely accepted  idea at that time, due to Fermi, that the radius of a hadron is determined by the pion cloud and therefore should be approximately universal. More generally it was argued that all matrix elements involving color neutral heavy quark systems  should  be suppressed, leading to a strong reduction of the cross section of $J/\psi$-nucleon interaction ($\propto 1/M_{J/\psi}^2$) and ``an unusual conclusion that the nucleon becomes transparent to hadrons built of heavy quarks'' \cite{Frankfurt:1975sy}. This was a clear break with the strong interaction picture with one soft scale, which was common before the discovery of the $J/\psi$. A perturbative model for the  interaction of hadrons via two-gluon exchange was applied to $J/\psi-N $ interaction by Gunion and Soper~\cite{Gunion:1976iy} who demonstrated that within their model, the reduction of the $J/\psi-N$ interaction is related to the small spacial size of the $J/\psi$ meson.  Arguments that the suppression should also be  present in the non-perturbative domain, were given in~\cite{Frankfurt:1985cv}, where it was argued that reduced $J/\psi (\psi^{\prime})$-nucleon cross section extracted from the photoproduction measurements using the vector dominance model  significantly underestimates the genuine $J/\psi-N$ and especially $\psi^{\prime}-N$ cross section.

An independent development was the discussion of the hard exclusive processes at large four-momentum transfer squared ($Q^2$) such as the ones used to measure the  nucleon form factors and large angle hadron-hadron scattering in the high $Q^2$ limit.  A. Mueller~\cite{Mueller82} suggested the use of exclusive processes off nuclei, namely large angle reaction $pA \to pp (A-1)$ to discriminate between the two mechanisms of elastic $pp$ scattering: Brodsky-Farrar mechanism of interaction in small size configurations governed by hard gluon exchanges which leads to the cross section at large c.m. angles scaling with energy according to a power law with the exponent given by the number of constituents involved in the reaction~\cite{Brodsky:1973kr,Brodsky:1974vy} and the Landshoff mechanism of the three-gluon exchange in t-channel with the Sudakov form factor suppression \cite{Mueller:1981sg}. While S. Brodsky~\cite{Brodsky82} made a prediction that the cross section of the $\pi A\to \pi p (A-1)$ process should be proportional to the number of protons in the target, a debate ensued, whether the minimal Fock space components highly localized in space, give the dominant contribution in the kinematic range studied experimentally, or, the process is dominated by the end point contributions corresponding to quark-gluon configurations of average size (for a  review see~\cite{Radyushkin:2004sr}). The interplay between these two mechanisms was explored recently for the case of nucleon electromagnetic form factors~\cite{Kivel:2010ns}, where it was argued that in the intermediate range of $Q^{2}\sim 5 - 15~\mbox{GeV}^{2}$, the soft rescattering contribution is characterized by a semi-hard scale $Q \Lambda$, where $\Lambda \approx 0.7 ~\mbox{GeV}$. In such a scenario, one would expect a delayed onset of CT. It was also pointed out that if these processes as well as quasi-elastic electron-nucleus scattering are studied in the kinematics, where at least one  hadron in the final state has relatively small momentum, the space-time evolution of the quark-gluon wave packets involved in the collision must be taken into account, which greatly reduces the CT effects~\cite{FLFS88}.

 This called for finding high-energy processes, which are dominated by the interaction of hadrons in small size configurations that could be legitimately calculated in perturbative QCD (pQCD) and are not affected by the space-time evolution of the small wave packets. A key observation was that, due to the  possibility of treating these configurations as frozen during the collision process, one can introduce a notion of the cross section of a small dipole configuration (say $q\bar q$) of transverse size $d$, scattering off a nucleon\,~\cite{Frankfurt:1993it, Blaettel:1993rd}. In the leading log approximation, this cross section is given by\,~\cite{Frankfurt:2000jm}
\be
\sigma^{inel}_{q\bar q N}(d,x)= {\pi^2\over 3} \alpha_s(Q^2_{eff}) d^2\left[x G_N(x,Q^2_{eff}) +{2\over 3} x S_N(x, Q^2_{eff})\right],
\label{eq:pdip}
\ee
\noindent
where $Q^2_{eff} = \lambda/d^2$, $\lambda = $ 4 - 10, $x = Q^2_{eff}/s$, with $s$ the invariant energy of the dipole-nucleon system, $S$ is the  sea-quark distribution for quarks making up the dipole and $G$ is its counterpart for gluons. The value of $\lambda$ was estimated from matching the dipole description with the leading log description of $\sigma_L(x,Q^2)$ \cite{Frankfurt:1995jw}. In contrast to the previous estimates, this also includes contributions from quark exchanges, which is important for the interactions  at intermediate energies. Note that Eq.~\ref{eq:pdip} predicts a rapid increase of the dipole-hadron cross section with increasing energy, due to the rapid growth of $xG_N(x,Q^2)$ at small $x$. For example for $Q^2$ scale of  $\sim 3~\mbox{GeV}^{2}~(40~\mbox{GeV}^{2})$ typical for the $J/\psi~(\Upsilon)$ photoproduction this leads to the cross section growing as $s^{0.4} (s^{0.8})$.
This is qualitatively different from the expectation of the two-gluon exchange model~\cite{Gunion:1976iy}, where the cross section does not depend on the incident energy.

First, we consider the simple case of high energy CT, where only two conditions are required: dominance of small size configurations and weakness of the $q\bar q - N $ interaction. Next we will consider a more complicated case of  CT in the intermediate energy regime, where it is masked, to a large extent, by the expansion effects.  New methods for the study of CT effects as well as future perspectives of the experimental studies are discussed in the third and final sections of the review.

\section{Discovery of high energy CT}
At high energies the CT phenomenon arises from the fact that, exclusive processes on a nucleus at high momentum transfer preferentially select color singlet, small transverse size configurations. This is also referred to as  ``squeezing'', and the small size configuration then moves with high momentum through the nucleus with reduced interaction. This is because in coherent processes, the interaction between the small transverse size configuration and the nucleon is strongly suppressed, as the gluon emission amplitudes arising from different quarks cancel. This suppression of the interaction is one of the essential ingredients needed to account for Bjorken scaling in deep-inelastic scattering (DIS) at small $x \le 10^{-2}$ where  longitudinal distances essential in $\gamma^*- N $ scattering,
$\sim 1/(2m_Nx)$,
  significantly exceeds the nucleon size~\cite{Frankfurt:1988pr}. Thus the discovery of Bjorken scaling in DIS can be considered as the first indirect evidence for CT at high energies.

To directly observe high energy CT, one needs to find a process, which selects small transverse size configurations in the projectile. One idea is to select a special final state, such as, diffraction of a pion into two high transverse momentum jets. Qualitatively, one expects in this case the transverse size to be $d\sim 1/p_t(jet)$, where $p_t$ is the momentum component transverse to the direction of the jet. Another idea is to select a small initial state, such as, diffraction of a longitudinally polarized virtual photon into a meson. In this case, the transverse separation, $d$, between $q$ and $\bar q$ in the wave function of $\gamma^*_L$ decreases as $d\propto 1/Q$. The pQCD results for these processes were first derived in~\cite{Frankfurt:1993it,Brodsky:1994kf}, with the proofs of the QCD factorization given for di-jet production \cite{Frankfurt:2000jm} and for meson production \cite{Collins:1996fb} (where in addition to production of vector mesons, a general case of meson production: $\gamma^{*}_{L} + N \to ``meson \, system" + ``baryon\, system" $ was considered).

Accordingly, the phenomenon of high energy CT can be formulated in the form of a factorization theorem - namely that the amplitude of the exclusive hard process $``hard\,\,  probe" + \,target \to ``hadron"\,  + \,  ``final \,\,state \,\, of \,\, the \,\, target"$,  which is dominated by the contribution of the small size configurations  can be written as a convolution of three factors: {\it(1)} the wave function of the small size configuration produced by the hard probe, {\it(2)} the matrix element of the hard interaction and {\it(3)} the generalized parton distribution (GPD) in the target $f(x_1, x_2, t, Q^2)$, which describes the transition of the target from the initial state $i$ to the final hadronic state when partons with light-cone fractions $x_1$ and $x_2$ are exchanged with the hard block, $H$, and the final state $f$  of fixed mass and fixed $t=(p_i - p_f)^2$  is produced (see sketch in Fig. 1). The CT in this case ensures that additional exchanges of partons between the upper and lower blocks are suppressed by powers of $Q^2$. The nucleus GPD could enter in the diagonal regime - coherent scattering, in the nucleus break up kinematics or if a less restrictive condition $m_f -m_A \le const $ is imposed. It is worth emphasizing here that the use  of a hard probe does not automatically guarantee the validity  of the factorization theorem. For example in the case of the exclusive production of mesons, factorization has so far only been proven for the case of longitudinally polarized photons~\cite{Collins:1996fb}, while for the transverse photons, only the suppression of the Sudakov form factor of the unfactorized component is demonstrated~\cite{Mankiewicz:1999tt}. 

\begin{figure}[tb]
\begin{center}
\begin{minipage}[t]{6 cm}
\centerline{\epsfig{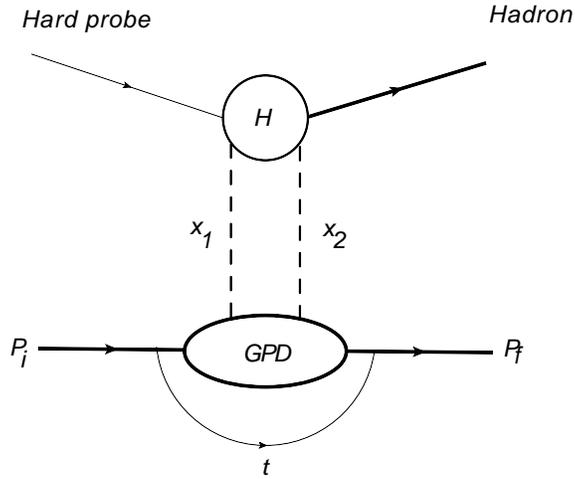}}
\end{minipage}
\begin{minipage}[t]{16.5 cm}
\caption{Sketch of the hard exclusive process in the factorization limit. Here $p_i (p_f)$ is the momentum of the initial (final) state, $t=(p_i - p_f)^2$, and $x_1$ and $x_2$ are the light-cone fractions of the exchanged partons.  \label{fig:sketch1}}
\end{minipage}
\end{center}
\end{figure}

\subsection{Pion dissociation into two jets
\label{sec:dijet}}
\begin{figure}[tb]
\begin{center}
\begin{minipage}[t]{5 cm}
\centerline{\epsfig{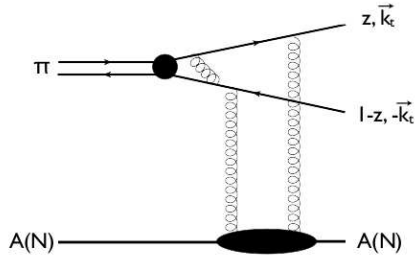}}
\end{minipage}
\begin{minipage}[t]{16.5 cm}
\caption{The two-gluon ladder exchange contribution to the pion's coherent diffractive dissociation process. \label{fig:dijet1}}
\end{minipage}
\end{center}
\end{figure}

Figure~\ref{fig:dijet1} shows  one of the dominant QCD diagrams for the coherent pion diffractive dissociation. The space-time picture of the process is as follows - long before the target, the pion fluctuates into $q\bar q$ configuration with transverse separation $d$, which elastically scatters off the target with an amplitude, which for $t = 0$ is given by Eq.(\ref{eq:pdip})  (up to small corrections due to different off shellness of the $q\bar q$ pair in the initial and final states), followed by the transformation of the pair into two jets. A slightly simplified final answer is
\be
A(\pi\,  N \to \, 2\, jets\, + \,N) (z,p_t,t=0) \propto \int d^2d \psi^{q\bar q}_{\pi}(z,d) \sigma_{q\bar q - N(A)}(d,s)e^{ip_td}, 
\label{eq:dijet}
\ee
where $z$ is the light-cone fraction of the pion momentum carried by a quark. The  normalization $\psi^{q\bar q}_{\pi}(z,d)$ -- the  quark-antiquark Fock component of the meson light-cone wave function,  at $d\to 0$ is determined by the Brodsky-Lepage relation \cite{Lepage:1980fj} 
\be
\psi^{q\bar q}_{\pi}(z,d)_{d\to 0}=\sqrt{48} f_{\pi} z(1-z),\ee
where $f_{\pi}=92$ MeV is the pion decay constant.
 
\begin{figure}[tb]
\begin{center}
\begin{minipage}[t]{8 cm}
\centerline{\epsfig{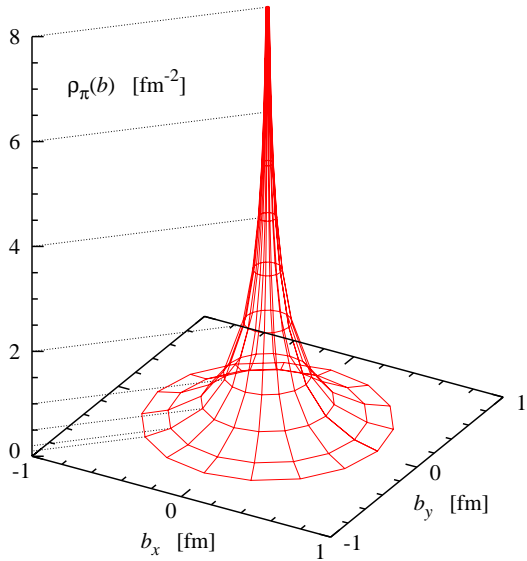}}
\end{minipage}
\begin{minipage}[t]{16.5 cm}
\caption{(Color online) Three--dimensional rendering of the transverse 
charge density in the pion based on the analysis of \cite{Miller:2010tz} .
\label{fig:rho3d}}
\end{minipage}
\end{center}
\end{figure}

 Note here that the presence of point-like configurations in the pion wave function is confirmed by the model independent analysis of the transverse pion charge density \cite{Miller:2010tz} , $\rho_{\pi}(b)$ which shows a sharp peak at $b\sim 0$ which appears to originate from the small size $q\bar q $ configurations (see Figure~\ref{fig:rho3d}).

\begin{figure}[tb]
\begin{center}
\begin{minipage}[t]{10 cm}
\centerline{\epsfig{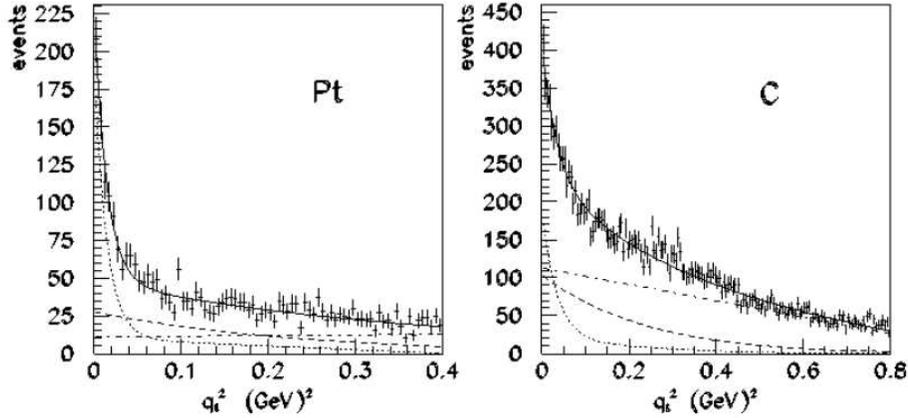}}
\end{minipage}
\begin{minipage}[t]{16.5 cm}
\caption{The Fermilab E791 di-jet yield from carbon and platinum as a function of the square of the transverse momentum transferred to the nucleus (from \cite{Aitala:2000hc}). The yields shown are for  $1.5\le k_t\le 2.0~ \mbox{GeV}$. The curves are Monte Carlo simulations  of the data: dotted line shows coherent dissociation, the dashed show incoherent dissociation, the background is shown by the dot-dashed line and the total by the solid line. \label{fig:dijet2}}
\end{minipage}
\end{center}
\end{figure}

\begin{figure}[tb]
\begin{center}
\begin{minipage}[t]{6 cm}
\centerline{\epsfig{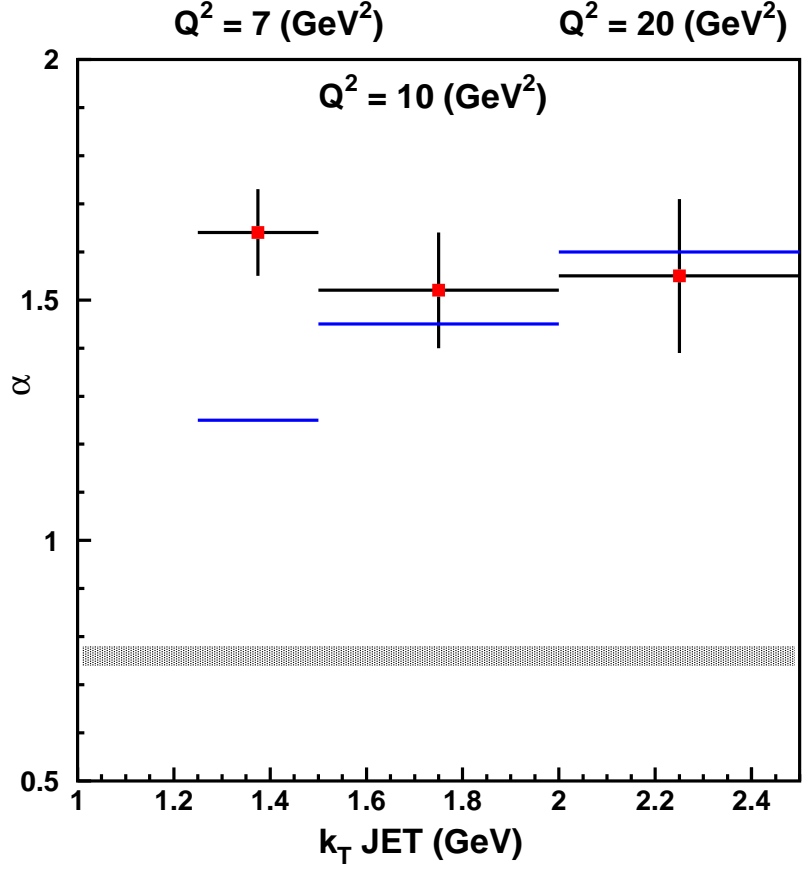}}
\end{minipage}
\begin{minipage}[t]{16.5 cm}
\caption{The values of $\alpha$ obtained from parametrization of the E791 di-jet cross section as  $\sigma = \sigma_{0} A^{\alpha}$. The data are shown as red points along with the quadrature sum of statistical and systematic errors and the $k_{T}$ bin size. The blue lines are the CT predictions of Ref.~\cite{Frankfurt:1993it} and the dark band is $\alpha \sim \frac{2}{3}$ observed in coherent inelastic diffractive pion-nucleus interactions. 
Typical virtualities $Q^2 =4k_t^2$ are also shown 
\label{fig:dijet3}.}
\end{minipage}
\end{center}
\end{figure}

The Fermilab experiment E791~\cite{Aitala:2000hc} measured the diffractive dissociation into di-jets of $500$ $\mbox{GeV}$ $\pi^{-}$ beam that coherently scattered from carbon and platinum targets. Diffractive di-jets were identified through the $e^{-bq^{2}_{t}}$ dependence of their yield, where $q^{2}_{t}$ is the square of the transverse momentum transferred to the nucleus and $b = \frac{<R>^{2}}{3}$, where $R$ is the nuclear radius. Figure~\ref{fig:dijet2} shows the $q^{2}_{t}$ distributions of di-jet events from platinum and carbon.  The events in the low-$q^{2}_{t}$ region are dominated by diffractive dissociation of the pion. The data are fit to sums of $q^2_t$ distributions of di-jet events produced coherently\footnote{Presence of the break up channel modifies the factor before the first exponential from $A^2$ to $(A-1)A$.} and incoherently from nuclear targets, and the background. The shapes of these distributions are calculated using Monte Carlo simulations as shown in Fig.~\ref{fig:dijet2} for transverse momentum, $1.5\le k_t\le 2.0 \, \mbox{GeV}$. The per-nucleon cross section for di-jet production is parametrized as $\sigma = \sigma_{0} A^{\alpha}$, where $\sigma_{0}$ is the free cross section. The values of the exponent $\alpha$ obtained from the experiment E791 are shown in Figure~\ref{fig:dijet3} along with the CT predictions of \cite{Frankfurt:1993it}. This is to be compared  with the value of  $\alpha \sim  2/3$ measured in the reaction 
$\pi A\to 3 \pi A$ \cite{Zielinski:1983ty}. Note here that the inelastic coherent soft diffraction off nuclei: $h+ A\to X A$ is due  to  the  fluctuations of the strength of hadron-nucleon interaction, ($\sigma_{tot}$), and is dominated by the fluctuations near the average value \cite{Frankfurt:1993qi}. The exponent $\alpha$ for the soft diffraction drops when $A$ and/or $\sigma_{tot}$ increases. It was suggested \cite{Kopeliovich:1981pz} that inelastic diffraction is dominated by scattering off the small size configuration. This regime however is reached only for $A\gg 1000$ \cite{Frankfurt:1993qi}. At the
lowest $k_t$ range, there is some discrepancy between the experiment and theory, which may be interpreted as a manifestation of nonperturbative effects.

Overall, these results confirm the following CT predictions of~\cite{Frankfurt:1993it} : a) a strong increase of the $\pi +A \to "two \, jets" + A$ cross section with A ($\sigma \propto A^{1.61\pm 0.08}$) as compared to the prediction\footnote{In QCD a naive expectation of CT that the amplitude is proportional to A is modified \, \cite{Frankfurt:1993it,Brodsky:1994kf} due to the leading twist gluon shadowing which should be present at  sufficiently small x. This effect is not important for the x range of the experiment \cite{Aitala:2000hc}.} $\sigma \propto  A^{1.54}$, b) the $z^2(1-z)^2$ dependence of the cross section, where $z$ is the fraction of energy carried by the jet, and c) the dependence of the cross section on the transverse momentum of each jet with respect to the beam axis ($k_t$) behaves as $d\sigma/d^2k_{t}^2 \propto 1/k^{8}_{t}$. This scaling behavior is observed at relatively modest $k_t \ge 1.5~\mbox{GeV}$ indicating an early onset 
of the scaling. This maybe due to the presence of the plane wave factor in 
the final state rather than the vector meson wave function like in the case of the vector meson production discussed in the following subsection. Note that the CT prediction for the A-dependence was a factor of seven different from the A-dependence for the soft diffraction. The observed pattern is qualitatively different from the expectations of \cite{Bertsch:1981py} that the cross section should be $\propto A^{1/3}$ and fall exponentially with $k_t$.

In an update to the original analysis~\cite{Ashery:2005wa}, a fit to the $z$ distribution of the di-jet production cross section using Gegenbauer polynomials for different ranges of $k_t$ was reported. For $1.25\le k_t\le 1.5 \, \mbox{GeV}$, higher order polynomials appear to be important. Since, CT is observed for this $k_t$ range, this indicates that ``squeezing'' or the preferential selection of the  small transverse size configurations in the projectile occurs already  before the leading term $(1-z)z$ dominates. Therefore, it can be claimed that color transparency is unambiguously observed in the diffractive dissociation of $500~\mbox{GeV}$ pions into di-jets when coherently scattering from carbon and platinum targets.

The amplitudes of the processes $ \pi + A \to ``two \, jets'' + A$  were  also considered in Ref.~\cite{Braun:2001,Braun:2002}. The authors evaluated amplitudes of the processes  
where the projectile pion  was substituted by the non-interacting $q\bar q$  pair. This approximation changes the kinematics of the process, in particular it changes  energy-momentum 
conservation.  Authors of Ref.~\cite{Braun:2002} found that this amplitude has a singularity in the physical domain of the process $q\bar q +A \to 2jet + A$  and stated that  as a result of this singularity color transparency phenomenon should be weakened. We want to emphasize that this  singularity is  absent in the amplitude of a physical process initiated by the 
projectile pion. Such a singularity contradicts the  Landau rules for the evaluation of singularities of matrix elements of the $S$ matrix of physical processes. Thus direct comparison 
of results obtained in Ref.~\cite{Braun:2001,Braun:2002} with the calculations of Ref.~\cite{Frankfurt:1993it} and data are not possible.  Note however that  results obtained 
in Ref.~\cite{Braun:2002} in the leading log approximation agree  with the result of Ref.~\cite{Frankfurt:1993it}. Another technical feature of 
Ref.~\cite{Braun:2001,Braun:2002} is that the distortion of kinematics described above requires also a  distortion of gauge identities. To elaborate  the role of 
these distortions of kinematics and gauge conditions the amplitude of a process $\pi + \gamma \to ``two \,\, jets'' $, where calculations are more simple,  was considered in Ref.~\cite{Frankfurt:2002gb}. The use of the gauge invariance in QED allows an unambiguous representation of the matrix element using the pion light-cone wave function~\cite{Frankfurt:2002gb}, leading to a result 
different from that in Ref.~\cite{Ivanov:2001ia} where the same approximations as in Ref.~\cite{Braun:2001,Braun:2002} were made.

\subsection{Photoproduction of $J/\psi$ 
\label{sec:jpsi}}
The $A$ dependence of $J/\psi$ production by real photons in the energy range of $80 - 190~\mbox{GeV}$ was studied on H, Be, Fe, and Pb targets at Fermilab \cite{Sokoloff:1986prl}. 
At these energies the leading twist gluon shadowing is still negligible, the coherence length is already large enough so the process  proceeds in three stages:
$\gamma$ converts into $ c\bar c $ pair before the target, with the size of the $c\bar c $ smaller than the average $J/\psi$ size (for large charm mass, $m_c$, the wave function 
of $J/\psi $ in the origin enters, see e.g.~\cite{Brodsky:1994kf}), the $c\bar c$ pair propagates through the target with little expansion and converts to $J/\psi$ outside the target.
 Accordingly, for these energies eikonal (higher twist) rescattering is expected to be small while the leading twist gluon shadowing is still negligible since $x_{eff} = M_{J/\psi}^2
/W^2 \sim 0.05$. Consequently, one expects proximity to the CT regime, which corresponds to the cross section of the sum of coherent and quasielastic process equal to 
\begin{equation}
{d\sigma (\gamma A \to J/\psi +A (A'))\over dt }
=(A(A-1) F_A^2(t) + A){d\sigma (\gamma N \to J/\psi +N)\over dt }.
\end{equation}
Here $F_A(t) $ is the nuclear form factor, $ F_A(t) \approx \exp(R_A^2t/6)$.
In the analysis of the experimental data, the contribution to the cross section proportional to $F_A^2(t)$ was studied and its integral over t was presented~\cite{Sokoloff:1986prl}. With complete  
color transparency the differential cross section for a nuclear target of radius $R_{A}$ will be of the form,
\be
\frac{d\sigma_{A}}{dt} = A^2\frac{d\sigma_{N}}{dt} e^{tR_{A}^{2}/3},
\ee

\noindent
where $d\sigma_{N}$ is the cross section for the nucleon target and corrections $\propto 1/A$ are neglected. From this, one can obtain the total cross section as 

\be
\sigma_A = \int dt \frac{d\sigma_{A}}{dt} \approx \frac{3A^2}{R_{A}^{2}}\frac{d\sigma_{N}}{dt} |_{t=0}.
\ee

The measured cross section can be parametrized as $\sigma_A = \sigma_1A^{\alpha}$, where $\sigma_1$ is a constant independent of $A$. One expects $\alpha = 4/3$ and the experiment measured $\alpha = 1.4 \pm 0.06 \pm 0.04$ for the coherently produced $J/\psi$. This result can be interpreted as due to CT effects at high energies. Nearly complete CT is also observed for quasi-elastic  (incoherent) $J/\psi$ production: $\alpha^{incoh}=0.94\pm 0.02\pm 0.03$.

\begin{figure}[htb]
\begin{center}
\begin{minipage}[t]{3 cm}
 \centerline{\epsfig{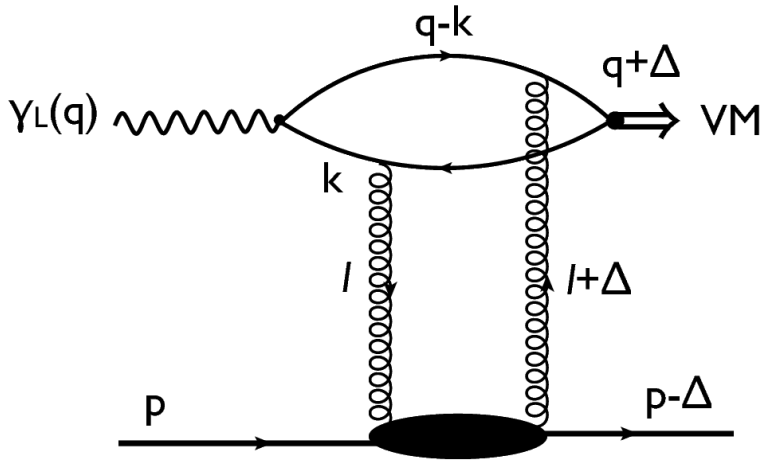}}
\end{minipage}
\begin{minipage}[t]{16.5 cm}
\vspace{-4 cm}
\caption{A two-gluon ladder  exchange contribution to vector meson production.  \label{fig:vm1}  
}
\end{minipage}
\vspace{-3.0 cm}
\end{center}
\end{figure}

\subsection{Vector meson production at HERA 
\label{sec:vmp}}

The leading twist picture (see Fig.~\ref{fig:vm1}) of exclusive vector meson production \cite{Brodsky:1994kf} is, in a sense, a mirror image of  the di-jet production. The longitudinally polarized virtual photon first transforms to a small transverse size pair, which interacts elastically with a target and next transforms to a vector meson. Hence the process is described by the same equation (\ref{eq:dijet}) as for the coherent pion scattering case with a substitution of the plane wave $q\bar q$ wave function by the $q\bar q$ wave function of the longitudinally polarized virtual photon. Some of the theoretical predictions of this leading twist picture include fast  $x$-dependence of the process at large $Q^2$, consistent with the $x$-dependence of the gluon distribution in the nucleon $G_N(x,Q_{eff}^2)$, and convergence of the $t$-dependence of the cross section to the universal one at large $Q^2_{eff}$, where it is given by the two-gluon form factor. 

\begin{figure}[tb]
\begin{center}
\begin{minipage}[t]{8 cm}
\centerline{\epsfig{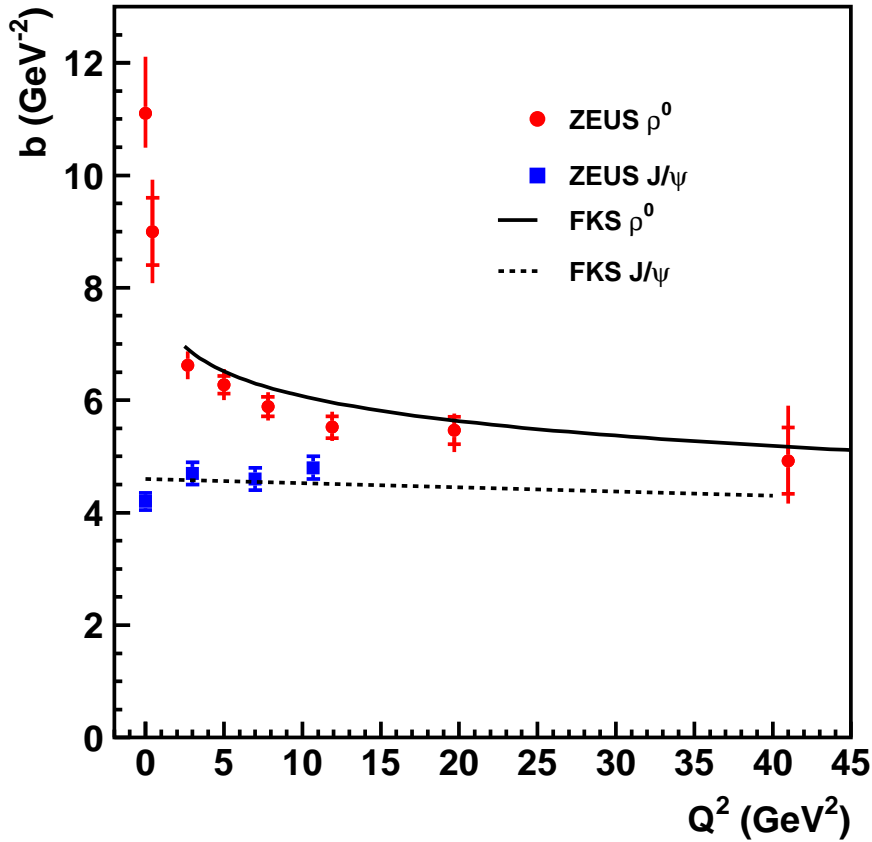}}
\end{minipage}
\begin{minipage}[t]{16.5 cm}
\caption{The convergence of the $t$-slopes of $\rho$ and $J/\psi$ electroproduction ($d\sigma/dt \propto exp(bt)$) at high $Q^2$, the data are from \cite{Chekanov:2004mw,Chekanov:2007zr}. The curves are predictions for the $Q^2$ dependence of $b$ from Ref.~\cite{Frankfurt:1997fj}. \label{fig:vm2}}
\end{minipage}
\end{center}
\end{figure}

Exclusive vector meson production was extensively studied at HERA, and the data  confirm these predictions~\cite{Frankfurt:2004vm}, where the convergence of the $t$-slope of $\rho$ and $J/\psi$ electroproduction at large $Q^2_{eff}$ is shown in Fig.~\ref{fig:vm2}. The data also confirm a conclusion of the model studies~\cite{Frankfurt:1995jw,Frankfurt:1997fj} that in a wide range of virtualities, one needs to take into account a higher twist effect of the finite transverse size of $\gamma^*_L$  to explain the absolute cross section and $t$-dependence of the data. The  leading twist dominance of the absolute cross section for all mesons and the $t$-dependence for light mesons requires very large $Q^2$ since only in this case one can neglect the transverse size of the $q\bar q$ pair in $\gamma^*_L$ as compared to that in the meson wave function. The same mechanism leads to $Q^2_{eff}/Q^2 \ll 1$  even at large  $Q^2$.

One of the open questions is the dynamics of the transverse cross section. On one hand, the wave function of $\gamma^*_T$ has a large size component, which could contribute to the vector meson production in the scattering of the transversely polarized photon, leading to a reduced CT in this channel. On the other hand, there is essentially no experimental evidence for a variety of large size effects, such as, larger t-slope for $\sigma_T$, and, an increase of $\sigma_L/\sigma_T$ as a function of $W$ for fixed $Q^2$. This suggests that a squeezing of the $\gamma^*_T$ wave function occurs at a nearly the same rate as for $\gamma^*_L$.   

To summarize this section. The presence of small size  $q\bar q$ Fock components in light mesons is unambiguously established. At  transverse  separations d $\le$  0.3 fm, pQCD reasonably describes small ``$q\bar q$ dipole''-nucleon interactions for $10^{-4} < x < 10^{-2}$. Color transparency is established for the small dipole interacting  with  the nuclei at $x \sim 10^{-2}$. Further  studies of high energy CT and onset of color opacity  will be performed  at the LHC in the  ultra-peripheral heavy ion collisions, see~\cite{Baltz:2007kq}  for a review.
There are also opportunities for probing CT in the processes with hadronic beams in the fixed target experiments with hadron momentum, $p_h \sim  100~\mbox{GeV}$ (see discussion in Section. \ref{sec:hadro-beam}).

\section{Color transparency for intermediate energies}
\subsection{Expansion effects}
In this section we discuss searches for CT at Jefferson Lab (JLab)~\cite{jlab} and Brookhaven National Lab (BNL)~\cite{bnl}, which correspond to the kinematics where the expansion or contraction of the interacting small size configuration is very important. This is because the essential longitudinal distances are not large enough to justify the use of the frozen approximation. And therefore it leads to strong suppression of the color transparency effect~\cite{FLFS88,Frankfurt:1988pr,jm90}. The maximal longitudinal  distance for  which coherence effects are still present is determined by the minimal characteristic internal excitation energies of the hadron. Estimates~\cite{FLFS88,Frankfurt:1988pr,jm90} show that for the case of the knock out of a  nucleon, the coherence is completely lost at distances $\mbox{l}_c \sim$ 0.4 - 0.6 $~\mbox{fm}\times  \, \mbox{p}_h$, with $\mbox{p}_h$ being measured in $\mbox{GeV}$. 
To obtain a large CT effect it is necessary that $l_c$ exceeds typical mean free 
path of a nucleon in nucleus - $r_{NN} \sim 2~fm$. Using our estimate of 
$l_c$ (Eq. \ref{lcoh}) we find that this corresponds to $p_N \ge 4~\mbox{GeV}$. For the $(e,e'p)$ reaction for the scattering off the nucleon with small momentum it corresponds to $Q^2\ge 2m_N E_N \sim 8~\mbox{GeV}^{2}$.

Quark-hadron duality suggests that similar distances are required for a quark to convert to a system of hadrons and start interacting strongly with a nuclear target. Indeed the analysis of the Giessen group~\cite{Giessen} suggests that the strength of interaction of the produced system at $z <  l_{c}$, where $z$ is the propagation distance, and the hadron formation time is similar to that for the formation of the hadron from a small color singlet as given by Eq.~\ref{eq:sigdif} discussed below\footnote{It is of interest that $l_{c}= 1 {\mbox{fm}} \times (p_h/M_h)$ assumed in most of the modeling of heavy ion collisions at the Relativistic Heavy Ion Collider (RHIC)~\cite{rhic} and the Large Hadron Collider (LHC)~\cite{lhc} which is much larger than the one given by Eq.~\ref{lcoh} for $M_h \le 1~\mbox{GeV}$}.
  
To describe the effect of the loss of coherence, two complementary languages were suggested. In Ref.~\cite{FLFS88} and based on the quark-gluon representation of the point-like configuration (PLC) \footnote{Note that the PLC is sometimes also referred to as the small size configuration, both terms are used interchangeably in the text.} of the hadron wave function, the time of decoherence was estimated using energy-time uncertainty relation for a fast nucleon: 
\begin{equation}
\Delta E= \sqrt {p_h^2 +m_{inter}^2} - \sqrt {p_h^2 +M_{h}^2}\approx
{\Delta M_h^2 \over 2p_h }.
\end{equation}
where $ \Delta M_h^2= m_{inter}^2- M_{h}^2$ is the typical energy  
non-conservation in the intermediate state. Hence  
 \begin{equation}
 l_{c}={2p_h\over \Delta M_h^2}. 
 \label{lcoh}
\end{equation}
 and $\Delta M_h^2 \sim 0. 7~\mbox{GeV}^2 $ based on the additive quark model wave function and $\sim 1~\mbox{GeV}^2 $ based on the slope of the Regge trajectories. It was argued based on the structure of the light-cone energy denominators that the expansion of the wave packet follows the quantum diffusion pattern such that~\cite{FLFS88}  
\begin{eqnarray}
\sigma^{PLC}(z) =(\sigma_{hard} + {z\over l_{c}}[\sigma 
-\sigma_{hard}])
\theta(l_{c}- z) +\sigma\theta\left(z-l_{c}\right), 
\label{eq:sigdif}\end{eqnarray}
where $\sigma$ is the hadron-nucleon cross section and $\sigma_{hard}$ is the transverse area occupied by quarks over which they convert into hadrons. This equation is justified for an early  stage of time development in the leading logarithmic approximation when pQCD can be applied. Also, one can expect that Eq.~(\ref{eq:sigdif}) smoothly  interpolates between the hard and soft regimes. A sudden change of $\sigma^{PLC}$ would be inconsistent with the observation of an early (relatively low $Q^2$) Bjorken scaling \,\cite{Frankfurt:1988pr}.  Eq.(\ref{eq:sigdif}) implicitly incorporates the geometric scaling for the PLC-nucleon interactions, which for the discussed energy range includes nonperturbative
effects.

The time development of the $PLC$ can also be obtained  using a baryonic basis for the PLC wave function:

\begin{eqnarray}
\displaystyle{\left| \Psi_{PLC}(t)\right>=\Sigma_{i=1}^{\infty} a_i \exp(-iE_it)\left| \Psi_{i} \right>
= \exp(-iE_1t)\Sigma_{i=1}^{\infty} a_i \exp\left({-i(m_i^2-m_1^2)t\over 2P}\right)\left| \Psi_{i} \right>},
\end{eqnarray}

where $\left| \Psi_{i} \right>$ are the eigenstates of the Hamiltonian with masses $m_i$, and $P$ is the momentum of the PLC, which satisfies 
$E_i \gg m_i$.  As soon as the relative phases of the different hadronic components  become large (of the order of one), the coherence is likely to be lost. Hence $l_{c}$ is actually the length at which coherence between the lowest and the first excited state is lost.

Numerical results of the quantum diffusion model \cite{FLFS88,Frankfurt:1988pr} and the model based on the expansion over hadronic basis with sufficiently large number of intermediate states~\cite{jm90} are pretty close. However, although both approaches model certain aspects of the dynamics of expansion, a complete treatment of this phenomenon in QCD is so far missing. In particular, the phenomenon of spontaneously broken chiral symmetry may lead to the presence of two scales in the expansion rate, one corresponding to the regime where quarks can be treated as mass-less, and another where virtualities become small enough and quarks acquire effective masses of the order of 300 MeV.

\subsection{Large angle quasielastic A(p,2p) process}
\begin{figure}[htb]
\begin{center}
\begin{minipage}[t]{7.5 cm}
\centerline{\epsfig{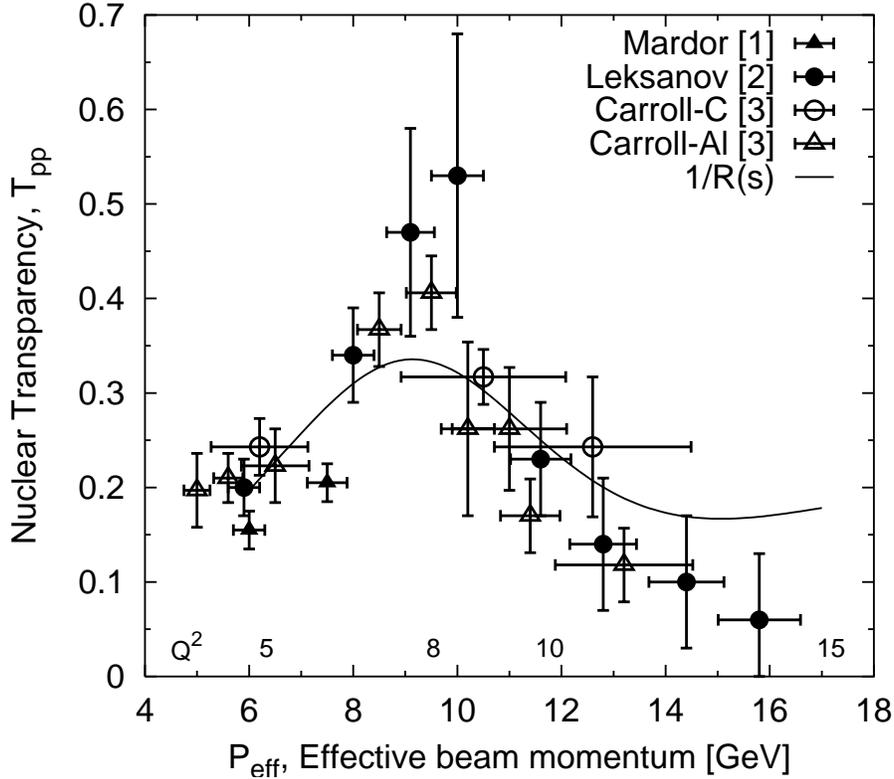}}
\end{minipage}
\begin{minipage}[t]{16.5 cm}
\caption{The nuclear transparency for $^{12}$C and $^{27}$Al (scaled by $(\frac{27}{12})^{1/3}$) versus the effective beam momentum. The curved line is the inverse of $R(s)$ as defined in the text~\cite{Aclander:2004zm}. \label{fig:Ap2p}}
\end{minipage}
\end{center}
\end{figure}

 The first attempt to measure the onset of CT at intermediate energies took place at BNL using the large angle $A(p,2p)$ reaction~\cite{Carroll:88}. In this experiment large angle $pp$ and quasi-elastic $(p,2p)$ scattering were simultaneously measured in hydrogen and several nuclear targets, at incident proton momenta of $6 - 12~\mbox{GeV}$. The nuclear transparency was measured as the ratio of the quasi-elastic cross section from a nuclear target to the free $pp$ elastic cross section. The transparency was found to increase by more than a factor of 2, consistent with the CT prediction, between $6 - 9.5~\mbox{GeV}$ but fell significantly between $9.5$ and $12~\mbox{GeV}$. This experiment was followed  by a dedicated experiment EVA~\cite{Leksanov:01}, which extended these measurements to $14.4~\mbox{GeV}$. Due to the Fermi motion, the invariant energy of $pp$ collision, $s_{pp}$ differs from that for the scattering off hydrogen, therefore an effective incident momentum was introduced and defined as $s_{pp} = 2m_p\sqrt{m_p^2+p^2_{eff}} + 2m_p^2$, it  ranges from $5.0 - 15.8 ~\mbox{GeV}$. The final results from both experiments~\cite{Aclander:2004zm} are shown in Fig~\ref{fig:Ap2p}. 
  The initial increase in transparency with energy followed by a decrease at higher energies was confirmed in the new experiment. In addition to the energy dependence of the transparency, the angular dependence (80 $< \theta_{c.m.} <$ 90$^\circ$) was also measured.


The initial rise in transparency between $\mbox{p}_p = 5.9$ and $9.5 ~\mbox{GeV}$ is consistent with the selection of a point like configuration and its subsequent contraction (for the initial proton) and expansion for final protons over distances comparable to Eq. \ref{lcoh} with $\Delta M^2 \sim 1\, \mbox{GeV}^2$.
 Calculations within the eikonal approximation  with proper normalization of the wave function agree well with the $\mbox{p}_p =5.9~\mbox{GeV}$ data. The transparency increases significantly for $\mbox{p}_p = 9~\mbox{GeV}$ where $l_{c} \sim 4 $ fm for the projectile proton. Hence, momenta of the incoming proton $\sim  10~\mbox{GeV}$ are  sufficient to rather significantly suppress expansion effects, which implies that one can use proton projectiles with  energies above $\sim 10~\mbox{GeV}$ to study other aspects of the strong interaction dynamics. At the same time, the  reported calculations, using the eikonal approximation, for $\mbox{p}_p = 11.5 - 14.2~\mbox{GeV}$ represents a problem for all current models including those which were specifically suggested to explain initial indications of the non-monotonous energy dependence of the transparency. 

Two possible explanations have been suggested for the observed drop in transparency for $\mbox{p}_p \ge 9~\mbox{GeV}$. One suggested that the energy dependence arises from an interference between two distinct amplitudes, that contribute to the $pp$ elastic scattering~\cite{Ralston:1988rb,Jain:1995dd}. One amplitude is a hard amplitude, which should dominate the high energy cross section while the other is an amplitude due to exchange of three gluons in the t-channel - know as the  Landshoff mechanism. It is suppressed by the Sudakov  form factors at large energies \cite{Mueller:1981sg} but may be significant at intermediate energies. The $pp$ elastic scattering cross section near $90^{\circ}_{c.m.}$ degrees varies with c.m. energy $(s)$ as
\begin{equation}
\frac{d\sigma}{dt_{pp}}(\theta = 90^{\circ}_{c.m.}) = R(s) s^{-10}
\end{equation}
It is assumed in the model that the coherence length is much larger than the estimate above so that the expansion of the small size configurations can be neglected. It is further assumed that the contribution of  the long-ranged contribution of the Landshoff mechanism is completely attenuated by the nuclear matter resulting in  the interference disappearing for the nuclear cross section and hence the energy dependence of the transparency should be the inverse of $R(s)$, as shown by the curve in Fig~\ref{fig:Ap2p}. This mechanism is called nuclear filtering. A recent relativistic multiple-scattering Glauber approximation (RMSGA) calculation~\cite{Overmeire:2006} which includes both CT and nuclear filtering 
provides an acceptable fit to most of the data points. 

The second explanation \cite{Brodsky:1987xw} suggests that the energy dependence of the $pp$ elastic scattering cross section scaled by $s^{-10}$ corresponds to a resonance or a threshold for a new scale of physics, such as charmed quark resonance or other exotic QCD multi-quark states.  However, because the drop of the transparency occurs over a large range of $s_{pp}$:    $ 24~\mbox{GeV}^{2}\le s_{pp}\le  30~\mbox{GeV}^{2}$, it is  too broad for a resonance \cite{Brodsky:1987xw}  or for interference of quark exchange and Landshoff mechanisms \cite{Ralston:1988rb,Jain:1995dd}\,. In any case the trend, if confirmed by future data at higher energies, would strongly suggest that the leading power  quark exchange mechanism of elastic scattering dominates in $pp$ scattering only at very large energies.
 
There  exist an independent evidence for the importance of the quark exchange, which comes from a systematic study of a large variety of reactions  for incident momentum between $6$ and $9.9~\mbox{GeV}$ and below \cite{White:1994tj}. It  has found that cross sections of the processes where quark exchanges are allowed are much larger, and the energy dependence is roughly consistent with quark counting rules. Among the biggest puzzles is the ratio of cross sections for $\bar p p $ and $pp$ elastic scattering at $\theta_{c.m.}=90^\circ$, which is found to be less than $0.04$ at $6~\mbox{GeV}$. At face value, it indicates extremely strong suppression of the diagrams with gluon exchanges in $t$ channel, though more systematic and more precise studies are clearly needed.

On the other hand  the  recent data from JLab studies of the large angle Compton scattering  are not described by  the minimal Fock space quark counting rule mechanism,  while they agree well  with predictions based on dominance  of the box diagram contribution~\cite{Radyushkin:2004sr,Danagoulian:2007gs}. This suggests that that even if the quark exchanges dominate at intermediate energies they are obtaining contributions both from short and large distance configurations in nucleons.
   
\subsection{Quasielastic Electron Scattering on Nuclei}
Compared to hadronic probes, the weaker electromagnetic probe samples the complete nuclear volume. The fundamental electron-proton scattering cross section is smoothly varying and is accurately known over a wide kinematic range and detailed knowledge of the nucleon energy and momentum distribution inside a variety of nuclei have been measured extensively in low energy experiments. Moreover, the energy transfer $\omega$ and the momentum transfer $\vec{q}$  in electron scattering experiments can be varied independently allowing the flexibility to choose kinematics regions that are relatively clean for investigating the onset of CT. The advantages of electro-nuclear reactions was immediately recognized following the BNL $(p,2p)$ experiments and efforts to measure CT using electron scattering were launched. In these experiments quasi-elastic (Bjorken $x \approx 1$) electrons scattering was used to tag the knocked-out proton from a nuclear target.

In quasi-elastic $(e, e'p)$ scattering from nuclei, the electron scatters from a single proton, which is moving due to its Fermi momentum~\cite{Frulani:1984}. The relatively high-energy resolution of the experiments allows clean detection of quasi-elastically scattered protons over a large kinematic range ensuring exclusivity. Additionally, the $(e,e'p)$ reaction under quasi-elastic conditions is relatively less sensitive to the unknown large momentum components of the nuclear wave function~\cite{Frankfurt:1988pr}. In the plane wave impulse approximation (PWIA) the proton is ejected without final state interactions with the residual (A-1) nucleons. At high momentum transfers, the measured $A(e, e'p)$ cross section would be reduced compared to the PWIA prediction in the presence of final state interactions, where the proton can scatter both elastically and inelastically from the surrounding nucleons as it exits the nucleus. The deviations from the simple PWIA expectation is used as a measure of the nuclear transparency. Note that processes where the proton scatters inelastically involve nuclear excitation energies well above the pion mass (typically of the order of $\Delta $ - N mass difference)  which are not included in the experimental cross section.  In the limit of complete color transparency, the final state interactions would vanish and the nuclear transparency would approach unity. Numerically nuclear transparency can be written as
\be
T(Q^2) = \frac{\int_{V}d^3p_mdE_m Y_{exp}(E_m,\vec{p}_m)}{\int_{V}d^3p_mdE_m Y_{PWIA}(E_m,\vec{p}_m)},
\label{transp}
\ee
where the integral is over the phase space V defined by the cuts on missing energy $E_m$ (typically $<$ 80 MeV) and missing momentum $|\vec{p}_m|$ (typically $<$ 300 MeV), $Y_{exp}(E_m ,\vec{p}_m)$ and $Y_{PWIA}(E_m ,\vec{p}_m)$ are the corresponding experimental and PWIA yields. The $E_m$ cut prevents inelastic contributions above pion production threshold. In the conventional nuclear physics picture one expects the nuclear transparency to show the same energy dependence as the energy dependence of the $NN$ cross section. Other effects such as short-range correlations and the density dependence of the $NN$ cross section will affect the absolute magnitude of the nuclear transparency but have little influence on the energy ($Q^2$) dependence of the transparency. Thus the onset of CT would manifest as a rise in the nuclear transparency as a function of increasing $Q^2$.

The $(e,e'p)$ reaction is expected to be simpler to understand than the $(p,pp)$ reaction. The two explanations proposed to account for the observed energy dependence of nuclear transparency in  $(p,pp)$ reactions, as discussed in the previous section, involve, the oscillatory $s$-dependence of the $pp$ interaction at high energies~\cite{Ralston:1988rb} or a proton-proton resonance at energies near the threshold for $c\bar{c}$ production~\cite{Brodsky:1987xw}. 
These effects are obviously not relevant for the  $(e, e'p)$ reaction at high momentum-transfer as the effective $s$ is always kept equal to $m_p^2$.
One also works at the ejected proton energies where $\sigma_{tot}(pp) $ is a very weak function of incident energy.

The first electron scattering experiment to look for the onset of CT was the NE-18 A(e,e'p) experiment at SLAC~\cite{Makins:1994}. This experiment
yielded distributions in missing energy and momentum completely
consistent with conventional nuclear physics and the extracted transparencies 
exclude sizable CT effects up to $Q^{2} = 6.8~\mbox{GeV}^2$ in contrast to the results from the A(p,2p) experiments~\cite{Carroll:88}. Later experiments with 
greatly improved statistic and systematic uncertainties compared to the NE-18 experiment~\cite{Makins:1994}, and with increased $Q^2$ range were carried out at JLab~\cite{Abbott:1998,Garrow:2002}.

\begin{figure}[h]
\begin{center}
\begin{minipage}[t]{6.0 cm}
\centerline{\epsfig{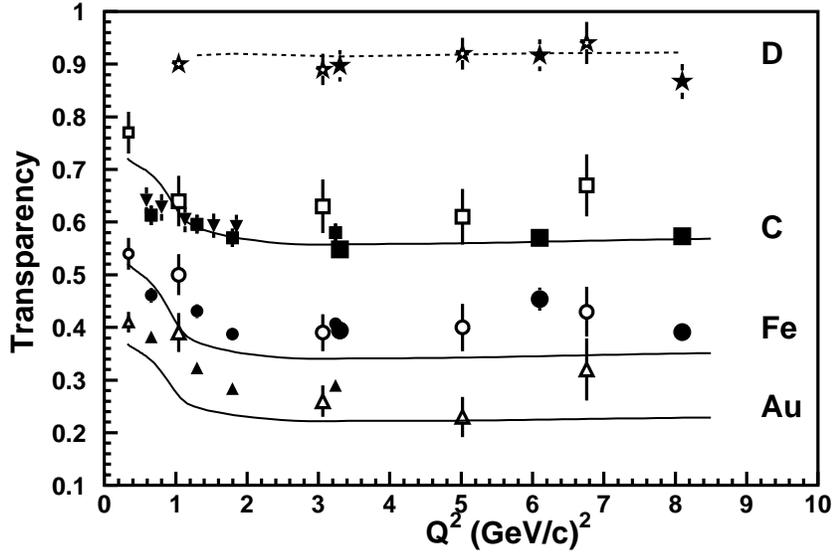}}
\end{minipage}
\begin{minipage}[t]{16.5 cm}
\caption{A compilation of transparency for $(e,e'p)$ quasielastic scattering 
from D (stars), C (squares, inverted triangles), Fe (circles), and Au (triangles) taken from Ref.~\cite{Garrow:2002}. Data from the three JLab experiments~\cite{Abbott:1998,Garrow:2002,Rohe:2005} are shown as solid points. The previous SLAC data~\cite{Makins:1994} are shown by large open symbols, and the previous Bates data~\cite{Garino:1992} are shown by small open symbols, at the lowest $Q^2$ on C, Ni, and Ta targets, respectively. 
The errors shown for the JLab measurement (solid points) include statistic and the point-to-point systematic (2.3\%) uncertainties, but do not include model dependent systematic uncertainties on the simulations or normalization-type errors. The net systematic errors, adding point-to-point, normalization-type and model-dependent errors in quadrature, are estimated to be (3.8\%), (4.6\%), and (6.2\%) corresponding to D, C, and Fe, respectively. The error bars for the
other data sets include their net systematic and statistical
errors. The solid curves shown from $0.5 < Q^2 <8.5~\mbox{GeV}^2$ are
Glauber calculations from Ref.~\cite{vijay:92}. In the case of D, the dashed
curve is a Glauber calculation from Ref.~\cite{Misak:95}.\label{fig:eep1}} 
\end{minipage}
\end{center}
\end{figure}

A compilation of the measured transparency $T(Q^2)$ values (defined as ratio of measured to PWIA  cross sections) from all electron scattering experiments are presented in Fig.~\ref{fig:eep1}. The errors shown include statistic and systematic uncertainties, but do not include model-dependent systematic uncertainties
in the spectral functions and correlation corrections used in the simulations for the JLab results (solid points). Data from all other experiments represented by open symbols include the full uncertainty.

The results show no $Q^2$ dependence in the nuclear transparency data 
above $Q^2 > 2~\mbox{GeV}^2$. The energy dependence below $Q^2 = 2 ~\mbox{GeV}^2$ is
consistent with the energy dependence of the $p$-nucleon cross section. 
Above $Q^2 = 2~\mbox{GeV}^2$,  excellent constant-value fits were obtained for the various transparency results. For deuterium, carbon, and iron these constant values were 0.904 ($\pm$0.013), 0.570 ($\pm$0.008),
and 0.403 ($\pm$0.008), with $\chi^2$ per degree of freedom of 
0.56, 1.29, and 1.17, respectively. In Fig.~\ref{fig:eep1} the measured transparency is compared with the results from correlated 
Glauber calculations, including rescattering through third order~\cite{vijay:92} (solid curves for $0.2 < Q^2 < 8.5~\mbox{GeV}^2$. In the 
case of deuterium the dashed curve shows a generalized eikonal approximation calculation~\cite{FGMSS:95} which coincides with a 
Glauber calculation for small missing momenta~\cite{Misak:95}. There are a number of Glauber-type transparency calculations for the $(e,e'p)$ reaction~\cite{Benhar:94,Lava:2004} 
which give similar results. Although these calculations can describe the $Q^2$ dependence of the 
nuclear transparencies, the absolute magnitude of the transparencies are under-predicted for the heavier nuclei. This behavior 
persists even after the model-dependent systematic uncertainties are accounted for. In Ref.~\cite{jain:93} it was suggested to treat absolute magnitude of the 
transparency relative to the measured value as a normalization factor $N(Q^2)$ 
-- which reflects fraction of configurations which are not filtered out --
and the strength of the absorption cross section for the interaction of these configurations  in nuclei - $\sigma_{eff}$ as two independent parameters. 
The lack of any $Q^2$ dependence of the A-dependence in the data and its consistency with Glauber model implies that within this approach $\sigma_{eff}$ is a constant with a value close to the free $NN$ cross section, implying that there is no squeezing.

All of the results reported by the various $(e,e'p)$ experiments were renormalized by a factor of 1.12 for carbon, 1.22 for iron and 1.26 for gold to 
account for shifts in the strength of the proton spectral function to higher missing energies and missing momentum, due to short range correlations.  However, the experiments were performed in the transverse kinematics where the missing momentum ($p_m$) has a very small longitudinal component
and wide range of the nuclear excitation energies. 
 Accordingly, the integral over the phase space volume in  Eq. \ref{transp}, written in terms of the proton momentum $k$ with longitudinal component $k_3$ and
 the  spectral function, $S_A(k, E_{rec})$ is proportional to 
\be
\int d E_{rec} d^3k  S_A(k, E_{rec}) \delta(k_3)  \propto \int d k k n_A(k),
\label{transint}
\ee
which is much less sensitive to the high momentum tail of the spectral function due to the short-range correlations than the full integral $\int S_A(k, E_{rec})d E_{rec}  d^3k$. As a  result  the  renormalization of the spectral function used in~\cite{Makins:1994,Abbott:1998,Garrow:2002} is not 
justified~\cite{Frankfurt:2000ty}. Thus we have replotted the transparency after removing these renormalization factors. 
\begin{figure}[h]
\begin{center}
\begin{minipage}[t]{4.0 cm}
\centerline{\epsfig{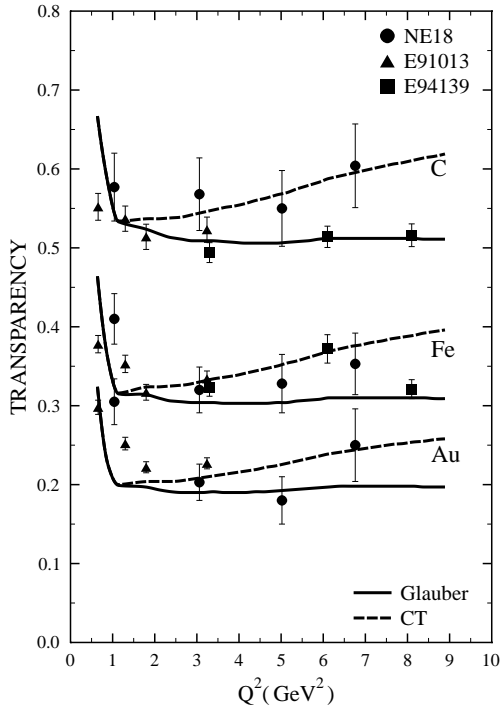}}
\end{minipage}
\begin{minipage}[t]{16.5 cm}
\caption{The same data as shown in Fig~\ref{fig:eep1}, but with the normalization due to SRC removed. The solid curves shown from $0.5 < Q^2 <8.5~\mbox{GeV}^2$ are Glauber calculations from Ref.~\cite{Frankfurt:2000ty}.\label{fig:eep1_2}}
\end{minipage}
\end{center}
\end{figure}
In Fig.~\ref{fig:eep1_2} the measured transparency with the renormalization removed is compared with the results from
Glauber model calculation~\cite{Frankfurt:2000ty}. These calculations use the Hartree-Fock-Skyrme nuclear 
spectral  function. It describes well the absolute value of the   (e,e') cross section  at $x = 1$,  and $Q^2 = 1\div 2~\mbox{GeV}^2$ 
which is proportional to the same integral as in  Eq.~\ref{transint}. These calculations are also in agreement with the differential cross sections of $(e,e'p)$ scattering measured in Ref.~\cite{Garrow:2002}. The dashed curves in the figure are the expectations of the CT model with $\Delta M^2=1~\mbox{GeV}^2$ (see Eq.~\ref{lcoh}) and squeezing starting at $Q^2= 1~\mbox{GeV}^2$: $\sigma_{hard} = \sigma_{NN}(1~\mbox{GeV}^2/Q^2)$. Note that after removing the renormalization factors, the calculated transparencies no longer under-predict the absolute magnitude of the measured transparencies  for heavier nuclei. 
\begin{figure}[htb]
\begin{center}
\begin{minipage}[t]{7 cm}
\centerline{\epsfig{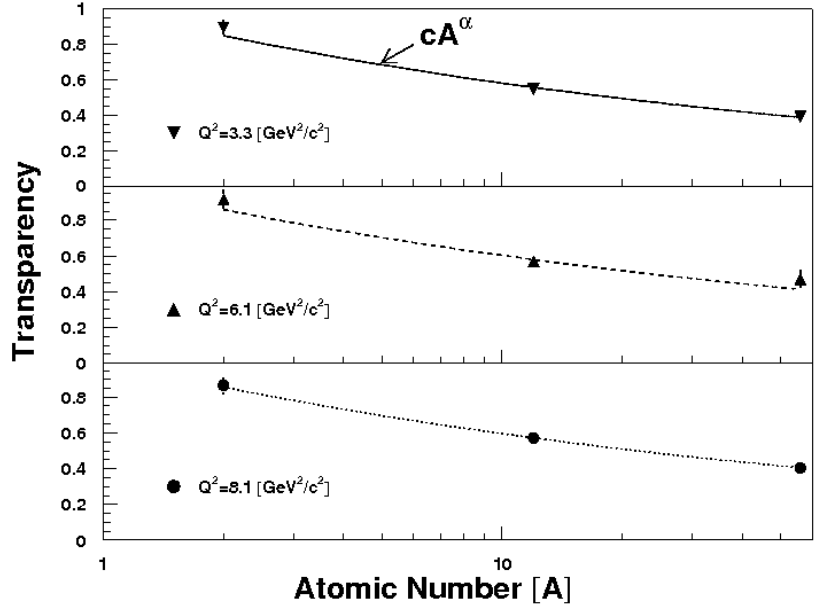}}
\end{minipage}
\begin{minipage}[t]{16.5 cm}
\caption{Nuclear transparency as a function of A at $Q^2 = 3.3, 6.1$,
and $8.1~\mbox{GeV}^2$ (top to bottom). The curves are fits to the D, C,
and Fe data using $T=cA^{\alpha}$ ~\cite{Garrow:2002}. \label{fig:eep2}}
\end{minipage}
\end{center}
\end{figure}

In addition to the  $Q^2$ dependence of the nuclear transparencies, the nuclear mass number $A$ dependence of the nuclear transparency was also studied by parametrizing the transparency to the form $T=cA^{\alpha(Q^2)}$. Empirical fits to this form for deuterium, carbon, and iron data are shown in Fig.~\ref{fig:eep2}. Within uncertainties, the constant $c$ is found to be consistent with unity as expected and the constant $\alpha$ to exhibit no $Q^2$ dependence up to $Q^2 = 8.1~\mbox{GeV}^2$ with a nearly constant value of $\alpha = -0.24$ for $Q^2 > 2.0~\mbox{GeV}^2$. This is also consistent with conventional nuclear physics calculations using Glauber approximation.

It was suggested in \cite{Frankfurt:1994nn} that the effect of CT can be compensated in an intermediate energy range by the effect of  suppression of point-like configurations in bound nucleons \cite{Frankfurt:1988pr} which maybe responsible for the suppression of nuclear cross sections at large x (the EMC effect).  The effect could be as large as 20\% for the cross section integrated over all nucleon momenta. However, this effect is proportional to the off-shellness of the nucleons which is much smaller in the 
 transverse kinematics, where $x=1$ and so the third component of the struck nucleon momentum is close to zero, and is expected to be $\sim$ 5 - 7\%.(see discussion after Eq.~\ref{transint} ).

The existing world data rule out any onset of CT effects larger than 7\% over the $Q^2$ range of $2.0 - 8.1~\mbox{GeV}^2$ (if one neglects the effect of the suppression of the point-like configurations in nuclei), with a confidence level of at least $90$\%. The $(e,e'p)$ data seem to suggest that a $Q^2$ of $8 ~\mbox{GeV}^2$ is not large enough to select the small transverse size objects in the hard $e-p$ scattering process.
 
\begin{figure}[htb]
\begin{center}
\begin{minipage}[t]{5 cm}
\centerline{\epsfig{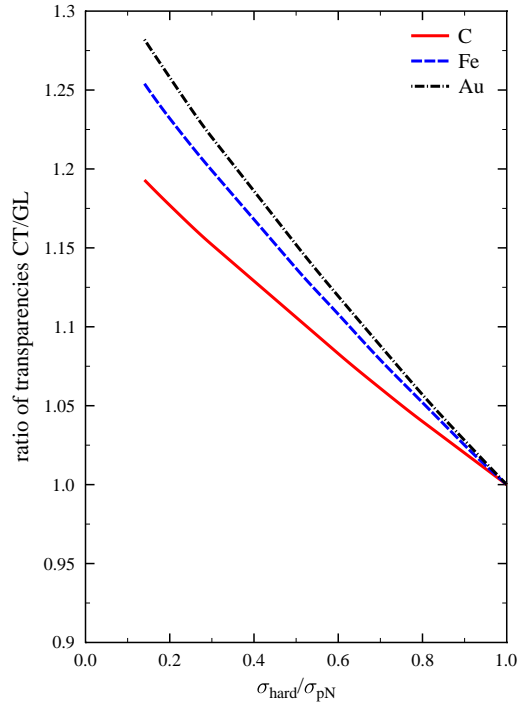}}
\end{minipage}
\begin{minipage}[t]{16.5 cm}
\caption{Ratio of the nuclear transparency calculated in the diffusion model (Eq.\ref{eq:sigdif}) and in the Glauber model as a function of $\sigma_{hard}/\sigma_{pN}$
for $Q^2= 8.1~\mbox{GeV}^2$ for Au, Fe, C (top to bottom). \label{T8}}
\end{minipage}
\end{center}
\end{figure}

 In order to quantify the constrains on the strength of the interaction of the produced quark-gluon system in the interaction point - $\sigma_{hard}$, we consider transparency for the highest $Q^2 $ of the current data  as a function of $\sigma_{hard}$ keeping the expansion rate consistent with the value fitting the EVA BNL data for $p_N\le 10~\mbox{GeV}$ ($\Delta M^2= 1~ \mbox{GeV}^2$). The results are shown in Fig. \ref{T8}. One can see from the figure that assuming that the increase of the transparency at $Q^2 = 8~\mbox{GeV}^2$ does not exceed 7\% we can exclude $\sigma_{hard}/\sigma_{pp}$ below $\sim 0.6~(0.4) $  if the point-like configuration suppression is neglected (included).

\subsection{Color transparency in meson production}

It is natural to expect that it is easier to reach  CT regime  for the interaction/production  of mesons than for baryons since  only two quarks have to come close together and a quark-antiquark pair is more likely to form a small size  object~\cite{Blaettel:1993rd}. Further, it is important to note that the unambiguous observation of the onset of CT is a critical precondition for the validity  of the QCD factorization theorem for exclusive meson production in DIS~\cite{Collins:1996fb}. This  is because where CT applies, the outgoing meson retains a small transverse size (inter-quark distance) while soft interactions like multiple gluon exchange between the meson produced from the hard interaction and the baryon are highly suppressed. QCD factorization is thus rigorously not possible without CT~\cite{gpdct}.

As described earlier, the $J/\psi $ coherent and quasielastic photoproduction experiments did find a weak absorption of $J/\psi$ indicating presence of CT. Support for CT was also observed in the coherent diffractive dissociation of $500~\mbox{GeV}$ negative pions into di-jets. There was also hints for CT in  several $\rho$-meson production experiments~\cite{Adams:1995,Airape:2003}, these are discussed in Sec.~\ref{sec:early_rho}. However all of these high energy experiments did not have good enough resolution in the missing mass to suppress hadron production in the nucleus vertex, making interpretation of these experiments somewhat ambiguous. Moreover, these high energy experiments do not tell us anything about the onset of CT. 

A recent high resolution experiment of pion production at JLab has reported 
evidence for the onset of CT~\cite{Clasie:2007} in the process $eA\to e\pi^+ A^*$. The chosen kinematics where $\vec{p}_{\pi}\|\vec {q}$ minimizes contribution of the elastic rescattering. The coherent length defined as the distance between the point where $\gamma^*$ converted to a $q\bar q$ and the interaction point - $l_{c}=2q_0/(Q^2 +M^2_{q\bar q})$ is small for the kinematics of \cite{Clasie:2007,Qian:2010} and varies weakly with $Q^2$. This simplifies  interpretation of the $Q^2$ dependence of the transparency as compared to the case of small x where  $l_{in}$ becomes comparable to the nucleus size. The experimental results  agree well with predictions of \cite{Larson:2006ge} and \cite{Cosyn:2008} where CT was calculated based on the quantum diffusion model - Eq. (\ref{eq:sigdif}). 

It is worth emphasizing, that  in the JLab kinematics one probes large x processes, which are dominated  for the pion case (and probably also for the $\rho$-meson case) in the pQCD limit by the contribution of the Efremov, Radyushkin, Brodsky, Lepage (ERBL) region -- knock out of $q\bar q$ pair from a nucleon. In this case $l_{in}$ has a different meaning than for small x processes where the DGLAP region dominates (emission of a gluon followed by the absorption of another gluon by the target). It corresponds to the longitudinal distance between the  point  where $\gamma^*$ knocks out a $q\bar q$ pair from the nucleon and the nucleon center.  This distance can be both positive and negative, and hence its variation does not lead to a change  of the rate of the absorption of the produced pair by the other nucleons. 
 
Results for the $\rho$-meson production from JLab also confirm the early onset of CT in mesons~\cite{ElFassi:2012nr}. To interpret this experiment one needs to take into account the effect of absorption due to decays of $\rho^0$ to two pions inside the nucleus, and the elastic rescattering contribution which is more important in this case than in the pion experiment since the data are integrated over a large range of the transverse momenta of the $\rho$ meson \, \cite{FMS07}. Up to these effects,  we  expect similar transparency for this reaction and for $\pi$-meson production. The pion electroproduction and rho experiments together conclusively demonstrate the onset of CT in the few $\mbox{GeV}$ energy range, these experiments are discussed in Secs.~\ref{sec:pion} and ~\ref{sec:rho}.

\subsubsection{Photo and Electroproduction of Pions }
\label{sec:pion}
{\it 1) Pion photoproduction}\\
~\\
The onset of CT in meson production was first explored in a pion photoproduction experiment at JLab. In this experiment nuclear transparency of the $\gamma n \rightarrow \pi^{-} p$ process was measured as a ratio of pion photoproduction cross section from $^4$He to $^2$H~\cite{Dutta:2003}. The $^4$He nucleus has several advantages as a choice for the studying the onset of CT. Exact nuclear ground state wave function are available for $^4$He~\cite{Arriga:96}, these along with the elementary hadron-nucleon cross-sections can be used to carry out precise calculations of the nuclear transparency~\cite{Gao:96}. Therefore, precise measurement of nuclear transparency from $^4$He nuclei  constitutes a benchmark test of traditional nuclear calculations. In addition, light nuclei such as $^4$He are predicted to be better for the onset of CT phenomenon because of their relatively small nuclear sizes, which are smaller than the length scales over which the hadrons of reduced transverse size revert back to their equilibrium size~\cite{FLFS88}. 

\begin{figure}[htb]
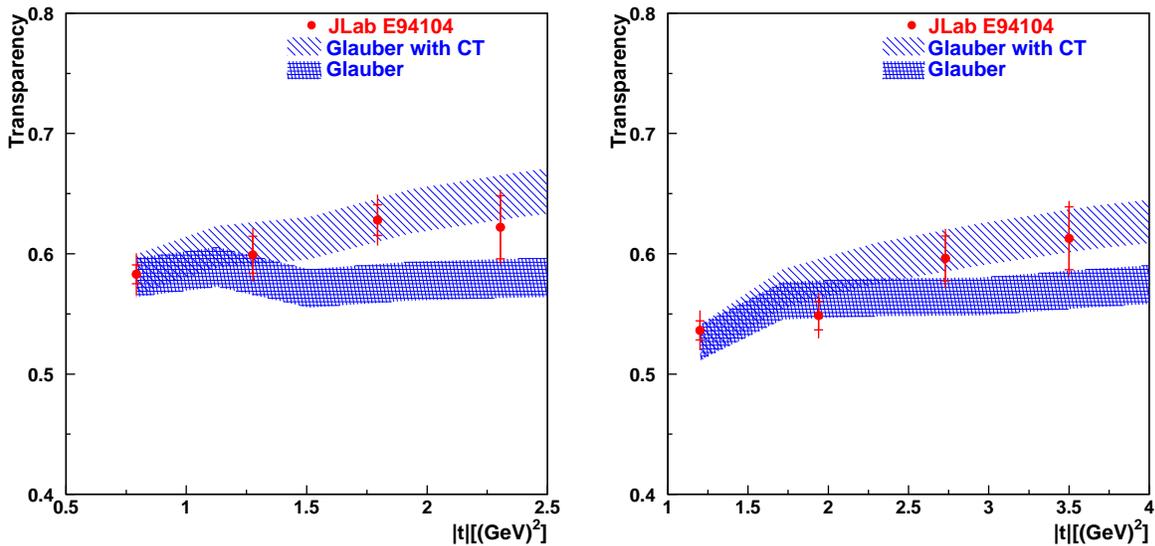

\begin{center}
\begin{minipage}[t]{5 cm}
\centerline{\epsfig{file=photopi1,scale=0.4}\epsfig{file=photopi2,scale=0.4}}
\end{minipage}
\begin{minipage}[t]{16.5 cm}
\caption{The nuclear transparency of $^4He(\gamma, p \pi^−)$ at
 $\theta_{cm}^{\pi} =70^{\circ}$ (left) and $\theta_{cm}^{\pi} =90^{\circ}$ (right), as a function of momentum transfer square $|t|$~\cite{Dutta:2003}. 
The inner error bars shown are statistical uncertainties only,
while the outer error bars are statistical and point-to-point
systematic uncertainties (2.7\%) added in quadrature. In addition
there is a 4\% normalization/scale systematic uncertainty
which leads to a total systematic uncertainty of 4.8\%.\label{fig:photopi}}
\end{minipage}
\end{center}
\end{figure}

The nuclear transparency extracted for the $^4$He target is shown in Fig.~\ref{fig:photopi}. Two calculations are compared with the measured transparency. One is a Glauber calculation which uses $^4$He configurations, which are snapshots of the positions of the nucleons in the nucleus, 
obtained from the variational wave function of Arriaga et al.~\cite{Arriga:96}. 
They contain correlations generated by the Argonne v14 and Urbana VIII models 
of the two-body and three-body nuclear forces respectively. The classical 
transparency was calculated from these configurations as described in~\cite{Gao:96}. The second calculation is also a Glauber calculation where the CT effect was included by modifying the hadron-nucleon total cross-section according to the quantum diffusion model~\cite{FLFS88}. The two calculations were normalized to each other at the lowest energy point. The momentum transfer squared $(|t|)$ dependence is not affected by the systematic uncertainties due to normalization and it is the $|t|$ dependence of the transparency that is of significance for the onset of CT phenomena.
The photopion results on $^4$He appears to deviate from the traditional nuclear physics calculations at the higher energies. The slopes
of the measured transparency obtained from the three points which are above 
the resonance region (above $E_{\gamma} = 2.25~\mbox{GeV}$ ) are in good agreement, within experimental uncertainties, with the slopes predicted by the calculations 
with CT and they seem to deviate from the slopes predicted by the 
Glauber calculations at the $ 1 \sigma (2 \sigma)$ level for $\theta_{CM}^{\pi}
= 70^{\circ} (90^{\circ})$. It is interesting that the deviation from Glauber calculations and the onset of scaling behavior in the cross-section for the $\gamma n \rightarrow \pi^{-} p$ and the $\gamma p \rightarrow \pi^{+} n$ processes~\cite{Zhu:2003} are observed at the same photon energies. Hence a change in the nuclear transparency may occur also due to transition from the vector meson dominated regime at small $t$ to the direct photon regime at $-t \ge 1.5~\mbox{GeV}^2$.

These data do suggest the onset of behavior predicted for CT, but future experiments with significantly
improved statistic and systematic precision are essential to confirm such conclusions. However, no evidence for CT was found when the data are
compared to a recent calculation~\cite{Cosyn:2006} in a relativistic and cross-section factorized framework, where the final state interactions of the ejected 
nucleon and pion are accounted for within a relativistic extension to the multiple-scattering Glauber approximation. Further progress will depend on the 
availability of new data.\\
~\\
\noindent
{\it 2) Pion electroproduction}\\
~\\
In 2004, the first extensive study of the pion electroproduction on a number of nuclear targets ($^1$H, $^2$H, $^{12}$C, $^{27}$Al, $^{63}$Cu and $^{197}$Au) was carried out at JLab, using the high-intensity, and continuous electron beams with energies up to $6~\mbox{GeV}$. This experiment (piCT) made it possible for the first time to determine simultaneously the $A$ and $Q^2$ dependence of the differential pion cross section for $Q^2 =1 - 5~\mbox{GeV}^2$~\cite{Clasie:2007,Qian:2010}. In the quasi-free approximation the pion electroproduction on a nucleus can be described as the incident electron exchanging a virtual photon with a proton which is moving inside the nucleus, the struck proton then ejects a positively charged pion and turns into a neutron. The ejected pion may interact with the residual nucleons and the fraction of pions which can escape from the nucleus is the pion nuclear transparency.  In the quasi-free picture, the ratio of the longitudinal to transverse cross section from a bound  proton inside the nucleus is expected to be the same as that from a free proton, this also provides the means to test the plausibility of the quasi-free approximation. Assuming the dominance of the quasi-free process, one can extract the nuclear transparency of the pions, by taking the ratio of the acceptance corrected cross sections from the nuclear target to those from the proton and/or deuteron. 
\begin{figure}[h]
\begin{center}
\begin{minipage}[t]{5.5 cm}
\centerline{\epsfig{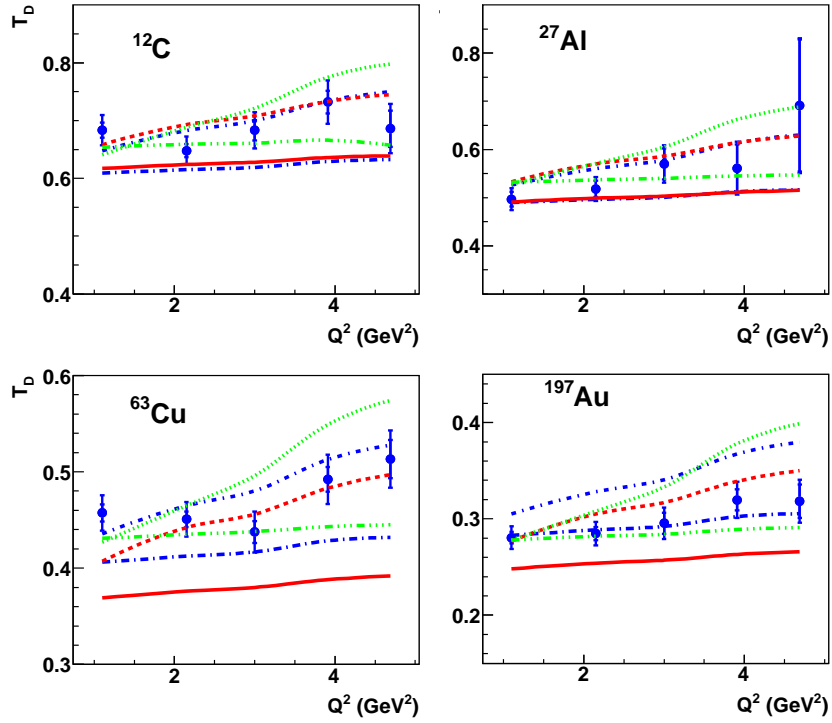}}
\end{minipage}
\begin{minipage}[t]{16.5 cm}
\caption{Nuclear transparency vs $Q^2$ for $^{12}$C, $^{27}$Al, $^{63}$Cu, 
and $^{197}$Au. The inner error bars are the statistical 
  uncertainties and the outer error bars are the statistical and
  point-to-point systematic uncertainties added in quadrature. The dashed and 
solid lines (red) are Glauber calculations  from Larson, {\it et al.}~\cite{Larson:2006ge}, with and  without CT, respectively. Similarly, the dot-short dash and dot-long  dash lines (blue) are Glauber calculations with and without CT from Cosyn, {\it et al.}~\cite{Cosyn:2008}. The dotted and dot-dot-dashed lines (green) are
  microscopic+ BUU transport calculations from Kaskulov {\it et al.}~\cite{mosel2}, with and without CT, respectively.~\label{fig:pict1}}
\end{minipage}
\end{center}
\end{figure}
The piCT experiment verified the dominance of the quasi-free process by comparing the ratios of the longitudinal to transverse cross sections from nuclear targets with those obtained from a nucleon target. Within experimental uncertainties, the $\sigma_L/\sigma_T$ ratios were found to be independent of $A$~\cite{Qian:2010}. This can be viewed as a confirmation of the quasi-free reaction mechanism. However, they cannot rule out non-quasi-free reaction mechanisms that affect the longitudinal and transverse character of pion electroproduction in a similar fashion. The experiment was intentionally restricted to $-t \le 0.5~\mbox{GeV}^2$ in order to minimize contributions from rescattering or multi-nucleon effects. 
Also, only pions emitted along the $\vec q$ direction were detected which practically eliminated contribution of the processes where the pion elastically rescattered in the final state.

In Ref.~\cite{Clasie:2007} the pion nuclear transparency was calculated as the ratio of pion electroproduction cross sections from the nuclear target to those from the proton. However, in order to reduce the uncertainty due the unknown elementary pion electroproduction off a neutron and uncertainties in the Fermi smearing corrections, the pion nuclear transparency was redefined in Ref.~\cite{Qian:2010} as the ratio of pion electroproduction cross sections from the nuclear target to those from the deuteron. Since the deuterium nuclear transparency is 
found to be independent of $P_{\pi}$ (or $Q^2$) with 81\% probability, both methods yielded almost identical $Q^2$ dependence of nuclear transparencies. The extracted transparency as a function of the pion momentum $Q^2$ for all targets is shown in Fig.~\ref{fig:pict1}.
 
The measured pion nuclear transparencies are compared to three different calculations. The calculations of  Larson, {\it et  al.}~\cite{Larson:2006ge}, use a semi-classical formula based on the eikonal approximation and a parametrization of the effects of final state interactions (FSI) 
in terms of an effective interaction. The effective interaction is based on the quantum diffusion model~\cite{FLFS88}, where the interaction of the small transverse size object PLC is approximately proportional to the propagation distance $z$ for $z<l_{c}$. In the limit of the coherence length $l_{c}=0$, a PLC is not created and the effective interaction reduces to a Glauber-type calculation with $\sigma_{eff} \approx \sigma_{\pi N}(P_{\pi})$. Cosyn {\it et al.} use a relativistic multiple-scattering Glauber approximation (RMSGA) integrated over the kinematic range of the experiment and compare it to a relativistic plane wave impulse approximation (RPWIA) to calculate the nuclear transparency~\cite{Cosyn:2008}. In RMSGA, the wave function of the spectator nucleon and the outgoing pion is taken to be a convolution of a relativistic plane wave and a Glauber-type eikonal phase operator that parametrizes the effects of FSI. CT was incorporated by replacing the total cross section with an effective one based on the quantum diffusion model~\cite{FLFS88}, similar to the effective interaction parameter of Larson, {\it et al.}~\cite{Larson:2006ge}. Finally, Kaskulov, {\it et al.}~\cite{mosel2,Mosel3} use a model built around a microscopic description~\cite{Mosel} of the elementary $^1$H($e,e'\pi^+$)n process, which is divided into a soft hadronic part and a hard partonic or deep inelastic scattering production part. For the reaction on nuclei, the elementary interaction is kept the same and nuclear effects such as Fermi motion, Pauli blocking and nuclear shadowing, are incorporated. Finally, all produced pre-hadrons and hadrons are propagated through the nuclear medium according to the Boltzmann-Uehling-Uhlenbeck (BUU) transport equation~\cite{mosel:2012}. The nuclear transparency is calculated as the ratio of the differential cross section calculated in this model, with and without FSI. The time development of the interactions of the pre-hadron is determined by the quantum diffusion model~\cite{FLFS88}. The production time and the formation time are taken from a Monte Carlo calculation based on the Lund fragmentation model~\cite{lund} as described in Ref.~\cite{galmc} which leads to similar parameters for expansion as the quantum diffusion model. Only the DIS part of the cross section is effected by the pre-hadronic interaction and thus in this model only the DIS events are responsible for the CT effect.
\begin{figure}[htb]
\begin{center}
\begin{minipage}[t]{7 cm}
\centerline{\epsfig{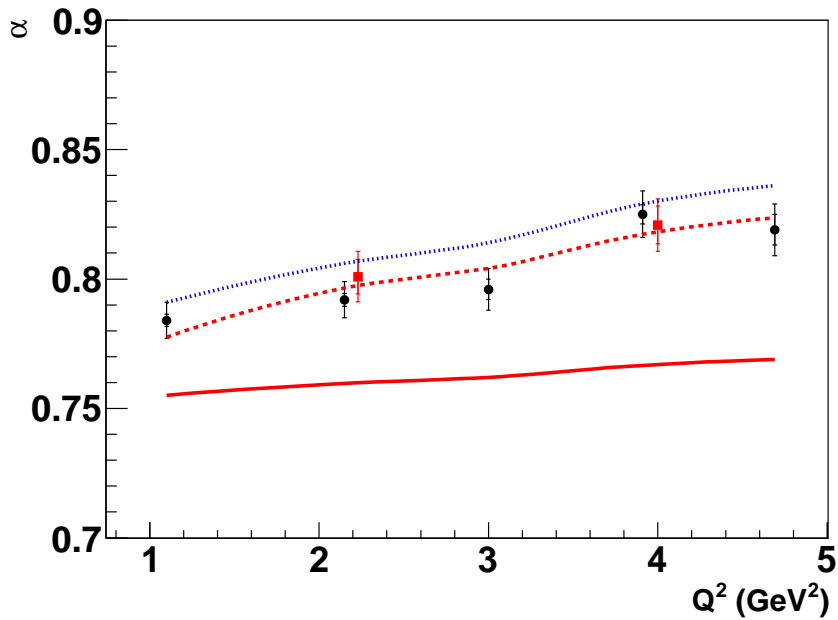}}
\end{minipage}
\begin{minipage}[t]{16.5 cm}
\caption{The parameter $\alpha (Q^2)$, as extracted from a fit of the nuclear transparency to the form $T=A^{\alpha - 1}$ (solid black circles and red squares). The inner error 
bars indicate the statistical uncertainties, and the outer error bars are the
    quadrature sum of statistical, systematic and modeling
    uncertainties. The solid,
    dashed and dotted lines are $\alpha$ obtained from fitting the
    $A$-dependence of the theoretical calculations: the Glauber and
    Glauber+CT calculations of Ref.~\cite{Larson:2006ge}, and
    the Glauber + CT (including short-range correlation effects)
    calculations of Ref.~\cite{Cosyn:2006}, respectively.
    The red squares show the $\alpha$ values extracted at the low $\epsilon$ kinematics.}  
    \label{fig:alpha}
\end{minipage}
\end{center}
\end{figure}

In the conventional nuclear physics picture the pion nuclear transparency is expected to be nearly constant over the pion momentum range of the experiment, because the hadron-nucleon cross sections are nearly independent of momentum over 
this range of momenta. Instead, the observed pion nuclear transparency results (as compared both to hydrogen and deuterium cross sections) show a steady rise versus pion momentum for the nuclear ($A>$ 2) targets, causing a deviation from calculations which do not include CT. The measured transparencies are in good agreement with the CT calculations of Larson, {\it et al.}, while the calculations of both Cosyn, {\it et al.} and Kaskulov, {\it et al.} overestimate the $P_{\pi}$ and $Q^2$ dependence of the data. However, it is more important to note that the rise in transparency in all the calculations that include CT are consistent with the measured rise in nuclear transparency versus $Q^2$, even though the underlying cause for the rise in nuclear transparency is different for the different model calculations.

The nuclear mass number $A$ dependence of the nuclear transparency gives further insight on the proper interpretation of the data in terms of an onset of CT. The entire nuclear transparency data set was examined using a single parameter fit to $T=A^{\alpha (Q^2) -1}$, where $A$ is the nuclear mass  number and $\alpha (Q^2)$  is the free parameter. Even though this single-parameter fit is simplistic and neglects local A-dependent shell or density effects, it does not
affect the final conclusion that the A-dependence changes with $Q^2$.
Thus, even though the exact value of $\alpha$ may come with a variety
of nuclear physics uncertainties, a significant empirical $Q^2$ dependence is observed from the data. In Fig. \ref{fig:alpha}, we compare $\alpha$ as function of $Q^2$, extracted from the single parameter form $T=A^{\alpha(Q^2) -1}$,  along with the calculations including CT effects of Larson, {\it et al.}~\cite{Larson:2006ge} and Cosyn, {\it et al.}~\cite{Cosyn:2006}. 
The results of the pion electroproduction experiment demonstrate that both the energy and $A$ dependence of the nuclear transparency show a significant deviation from the expectations of conventional nuclear physics and are consistent with calculations that include CT. The results can be seen as a clear indication of the onset of CT for pions.


\subsubsection{Early $\rho^0$  Meson Electroproduction Experiments}
\label{sec:early_rho}
\begin{figure}[htb]
\begin{center}
\begin{minipage}[t]{7 cm}
\centerline{\epsfig{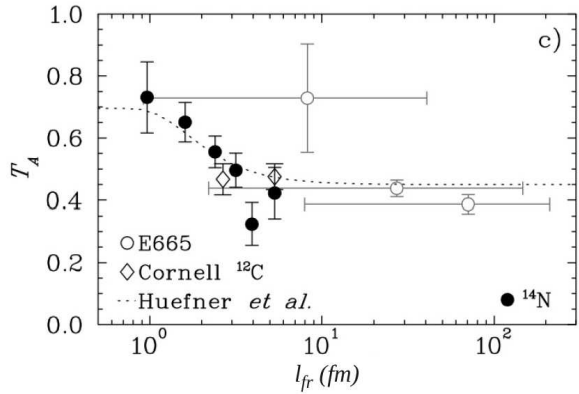}}
\end{minipage}
\begin{minipage}[t]{16.5 cm}
\caption
    {Nuclear transparency $T_A$ as a function of $l_{fr}$ for $^{14}$N (filled circle) targets from HERMES with previous experiments with photon (open diamonds) \cite{Mcclellan:1969uu} and muon (open circle) \cite{Adams:1994bw} beams. The dashed curves are the 
    calculation of H\"ufner {\it et al.} \cite{Hufner:1996dr} 
    within the eikonal approximation \cite{Bauer:1977iq}.}  
    \label{fig:lc-hermes}
\end{minipage}
\end{center}
\end{figure}
Electroproduction of vector mesons from nuclei is an excellent tool to investigate the formation and propagation of quark-antiquark ($q\bar{q}$) pairs under well-controlled kinematical conditions. This is due to the hadronic $q\bar{q}$ structure of high-energy photons \cite{Bauer:1977iq, Francis:1976sq, Shambroom:1982qj, Aubert:1985nj, Ashman:1987hu, Amaudruz:1991cc} arising from the virtual photon fluctuations to short-lived quark-antiquark states. These $q\bar{q}$ states of mass M$_{q\bar{q}}$ can propagate over a distance $l_{fr}$  -- photon coherence length -- without expansion (the frozen approximation)\footnote{Sometimes this is referred to as the coherence length, however, in line with the discussion in section 2 we will define coherence length  as the distance of evolution to the initial state to the hadron.}. It is given by the energy denominator for  the $\gamma^*\to 
 q\bar{q}$ transition 
 \be l_{fr}=2\nu/(Q^{2}+M_{q\bar{q}}^2)\approx {1\over xm_N}, \label{fr} \ee
 where $-Q^2$ and $\nu$ are the squared mass and energy of the photon in the lab frame (for reviews and references see e.g \cite{Bauer:1977iq, Piller:1999wx}).  In Eq.\ref{fr} at the last step we wrote answer in the Bjorken limit  and took into account that in the leading twist approximation for the vector meson production  $Q^2\gg M_{q\bar{q}}^2\gg m_V^2$.
 In exclusive diffractive production of the $\rho^0$ meson, which has the same quantum numbers as the virtual photon, a $q\bar{q}$ pair is scattered onto the physical $\rho^0$ mass shell by a diffractive interaction with the target \cite{Brodsky:1994kf, Gottfried:1969cd, Hufner:1996dr}. The HERMES collaboration at DESY \cite{Ackerstaff:1998wt} used exclusive incoherent electroproduction off $^1$H and $^{14}$N to study the interaction of the $q\bar{q}$ fluctuation with the nuclear medium by measuring the nuclear transparencies of  $^{14}$N relative to $^1$H as a function of the photon coherence length $l_{fr}$. The data were integrated over the excitation energy $\Delta E \le 0.6~\mbox{GeV}$ so production of $\Delta$-isobar was allowed.  Figure \ref{fig:lc-hermes} shows that the nuclear transparency for $^{14}$N  is decreasing as the photon coherence length increases. This decrease, known as the coherence length effect, is consistent with the onset of hadronic initial state interactions where the $q\bar{q}$ pair interacts with the nuclear medium like a $\rho^0$ meson. When the photon coherence length $l_{fr}$ is smaller than the mean free path of the $\rho^0$ meson in the nuclear medium, it is expected that
virtual photon converts into a  $q\bar{q}$ pair very close to the interaction point $\vec{r}$ so no absorption occurs before this point. Thus in  this limit   the nuclear transparency is independent of $l_{fr}$. The probability of the $q\bar{q}$ pair to interact with the nuclear medium increases with $l_{fr}$ until $l_{fr}$ exceeds the nuclear size \cite{Hufner:1996dr}. The dashed curve in Fig.  \ref{fig:lc-hermes} shows a theoretical prediction~\cite{Hufner:1996dr} calculated within the vector meson dominance model in the eikonal approximation \cite{Bauer:1977iq} neglecting non-diagonal transitions ($\rho\to \rho',...$).
 Its agreement with data suggests that the production process is dominated, given the relatively low $Q^2$ involved, by hadronic fluctuations which interact about as strongly as the produced $\rho^0$ meson.

The HERMES measurements have important implications for the study of color transparency using $\rho^0$ meson electroproduction, where the CT signal would be the increase of the nuclear transparency with $Q^2$, which controls the initial size of the $\rho^0$ meson. These results demonstrate that the increase of the nuclear transparency when $l_{fr}$ decreases ($Q^2$ increases) can mimic the CT effects. Therefore, to unambiguously identify the CT signal, one should keep $l_{fr}$ fixed while measuring the $Q^2$ dependence of the nuclear transparency, or perform the measurements in the regions where no $l_{fr}$ dependence is expected.
\begin{figure}[htb]
\begin{center}
\begin{minipage}[t]{7 cm}
\centerline{\epsfig{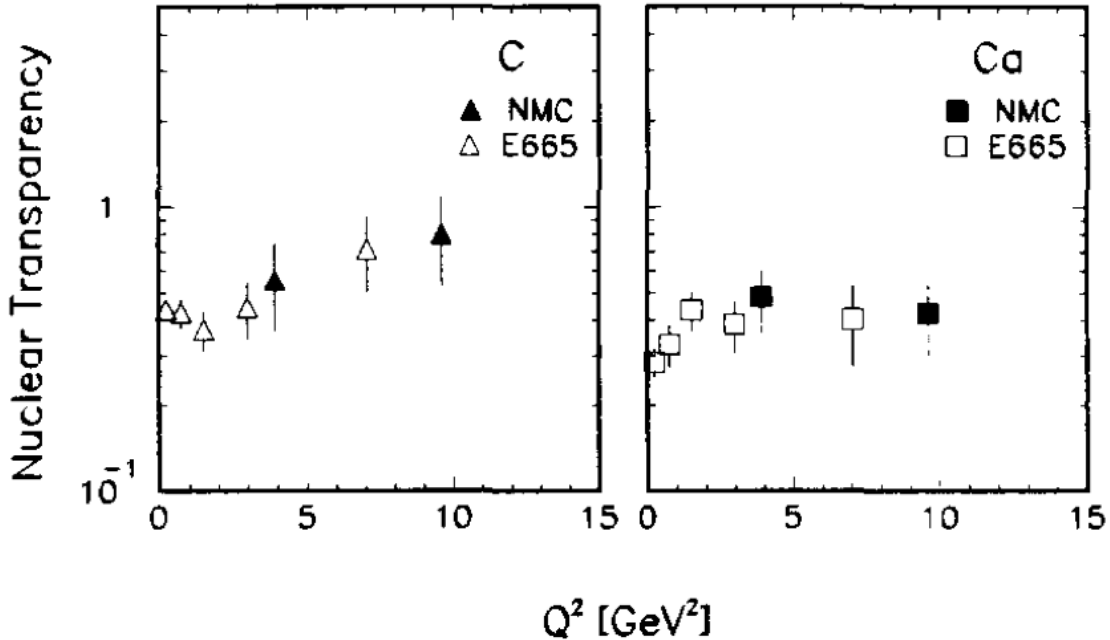}}
\end{minipage}
\begin{minipage}[t]{16.5 cm}
\caption
    {The $Q^2$ dependence of the nuclear transparency for incoherent exclusive $\rho^0$ virtual photoproduction. The data are from NMC \cite{Arneodo:1994id} (full symbols) and from E665 \cite{Adams:1994bw} (open symbols).}  
    \label{fig:nmc-e665-ct}
\end{minipage}
\end{center}
\end{figure}
When CT effects are present, a photon of high virtuality $Q^2$ is expected to produce a $q\bar{q}$ pair with small 
transverse separation which in the case of the longitudinally polarized photons 
will have reduced interaction in the nuclear medium. The dynamical evolution of this small size colorless $q\bar{q}$ pair to a normal size $\rho^0$ is controlled by the time scale called 
coherence time 
$t_{c}$. It  corresponds to the coherence  length $l_{c} = t_{c}$
given by Eq. \ref{lcoh} which in this case reads as 
\be l_{c} = 2\nu/({m_{v'}}^2 - {m_{v}}^2),\ee where $m_v$ is the mass of the $\rho^0$ in the ground state and $m_{v'}$ is the typical mass for the lowest $\rho$-meson excited states: $1.2 - 1.5~\mbox{GeV}$. 
The first measurements to study CT effects using incoherent diffractive $\rho^0$ leptoproduction off nuclei were performed at Fermilab by the E665 collaboration \cite{Adams:1994bw} and CERN by the NMC collaboration \cite{Arneodo:1994id}. Both experiments used muon beams with $450 ~\mbox{GeV}$ and $200~\mbox{GeV}$ energy, respectively. 
At these high energies, 
$l_{c}$ becomes larger than the nuclear diameter 
while $l_{fr}$ is comparable to the nuclear radius.
Therefore coherence length effects are not expected to play 
a major 
role in the CT signal because the fluctuations of the transverse size of the $q\bar{q}$ pair are mostly ``frozen" during the propagation. The two measurements are consistent with each other as shown in Fig \ref{fig:nmc-e665-ct}. While the NMC experiment reaches $Q^2$ values close to $10 ~\mbox{GeV}^2$, the measured nuclear transparency shows no $Q^2$ dependence. The E665 experiment measures an increase of the nuclear transparency with $Q^2$. This increase, however, is only suggestive of CT effects because of the limited statistical precision 
and need to subtract a substantial inelastic DIS contribution which was done using a Lund model based Monte Carlo. In addition, E665 performed $Q^2$-dependent fits to the exclusive $\rho^0$ incoherent production cross section, $\sigma_A$, from hydrogen, deuterium, carbon, calcium and lead by a power law $\sigma_A = \sigma_0 A^\alpha$ in which $\sigma_0$ and $\alpha$ are parameters. At low $Q^2$, the value of $\alpha$ measured is compatible with 2/3, a value characteristic of soft nuclear interactions. The observed increase with $Q^2$ shown in Fig. \ref{fig:e665-alpha} can only be indicative of CT effects due to the large statistical uncertainty associated with the highest $Q^2$ measurement.
\begin{figure}[htb]
\begin{center}
\begin{minipage}[t]{7 cm}
\centerline{\epsfig{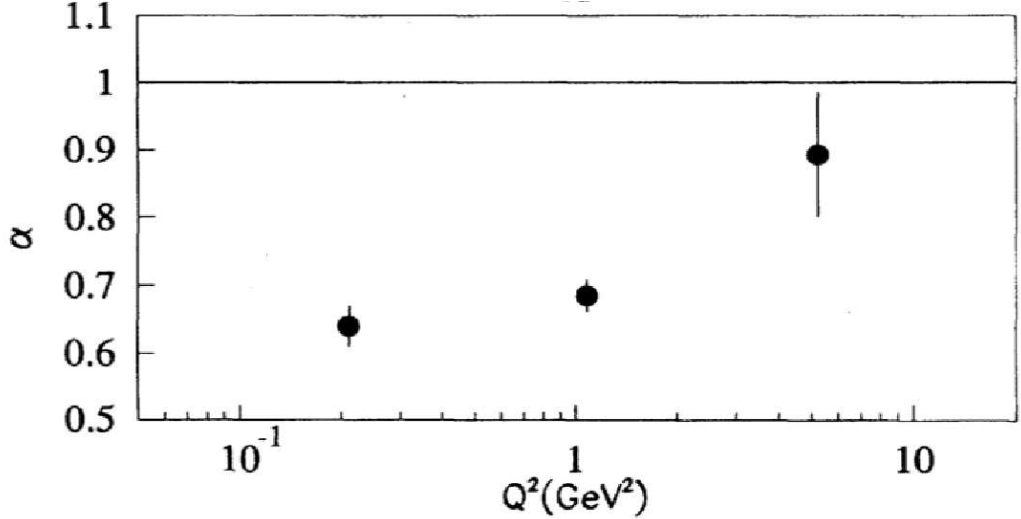}}
\end{minipage}
\begin{minipage}[t]{16.5 cm}
\caption
    {The parameter $\alpha$ as a function of $Q^2$ 
    for exclusive incoherent $\rho$-meson production.
   $\alpha$ = 1 corresponds to complete transparency. The errors shown in this plot are statistical only \cite{Adams:1994bw}.}  
    \label{fig:e665-alpha}
\end{minipage}
\end{center}
\end{figure}
The HERMES collaboration realized early on that in order to study CT effects through exclusive diffractive electroproduction of $\rho^0$ mesons, one needs to carefully disentangle coherence (shadowing) and formation (absorption) length effects. In addition to their measurements of the nuclear transparency as a function of the photon coherence length $l_{fr}$ (it was defined using Eq. \ref{fr} with $M^2_{q\bar{q}}= m^2_{\rho}$) for incoherent electroproduction of $\rho^0$ mesons off a nucleon in a nucleus, they did similar measurements for coherent $\rho^0$ production on a nucleus as a whole \cite{Airapetian:2002eh}. The nuclear transparency for coherent production was found to increase with the photon coherence length, as expected from the effects of the nuclear form factor \cite{Kopeliovich:2001xj}. Later, the nuclear transparencies for coherent and incoherent  $\rho^0$ production off $^1$H and $^{14}$N targets were used to study CT effects. And since these measurements strongly depend on $l_{c}$ because they are in the region where $l_{fr}$ is comparable to the nuclear radius, one has to study the $Q^2$ dependence of the nuclear transparency while keeping $l_{fr}$ fixed \cite{Airapetian:2002eh}. The HERMES experiment used $27.5 ~\mbox{GeV}$ positron beam off hydrogen and nitrogen targets. The scattered electron and two oppositely charged pions were required for the $\rho^0$ events. In addition, a cut on the two pions invariant mass $(0.6 < M_{\pi^+ \pi^-} < 1~\mbox{GeV}$ was applied to identify the $\rho^0$ meson and an exclusivity cut to the missing energy was used to favor the exclusive production. The $t' = t - t_{min} $ distributions were measured and the change in the slopes was a good indication to the transition from coherent to incoherent production. Cuts on $|t'| <  0.045~\mbox{GeV}^2$ for nitrogen and $|t'| <  0.4~\mbox{GeV}^2$ for hydrogen were used to identify coherent $\rho^0$ production, while for incoherent production, the 0.09 $< |t'| < 0.4 ~\mbox{GeV}^2$ cut was used for both targets. The nuclear transparencies were extracted in each ($l_{fr}$, $Q^2$) bin, and are shown in Fig.~\ref{fig:hermes-trans}. Due to the low statistics, the data have been fitted with a common $Q^2$-slope $(P_1)$, which has been extracted assuming $T_{c(inc)} = \sigma^{^{14}N}_{c(inc)}(l_{fr},Q^{2}) / A \sigma_{p} = P_0 + P_1\cdot Q^{2}$, letting $P_0$ vary independently in each $l_{fr}$ bin and keeping $P_1$ as common free parameter. The results are displayed as the lines in Fig.~\ref{fig:hermes-trans}. The common slope parameter of the $Q^2$-dependence, $P_1$, has been considered as  the signature of the CT effect averaged over the coherence length range. The $Q^2$ slopes for coherent and incoherent productions were found to be ($0.070 \pm 0.021$) and $(0.089 \pm 0.046)~\mbox{GeV}^{-2}$, respectively in agreement with existing model calculations \cite{Kopeliovich:2001xj}\footnote{A critical test of the CT interpretation of the data would be an observation of a significant reduction of the t-slope of the elementary cross section in the $Q^2$ range of the HERMES experiment. Unfortunately, data of sufficient accuracy is not available.}. These calculations are based on quantum mechanical description of the small size $q\bar{q}$ pair using the light-cone Green function formalism, which incorporate the effects of both coherence length and CT. These slopes were found to be positive and as such consistent with CT effects, however, the statistical significance of these results remains limited.
\begin{figure}[htb]
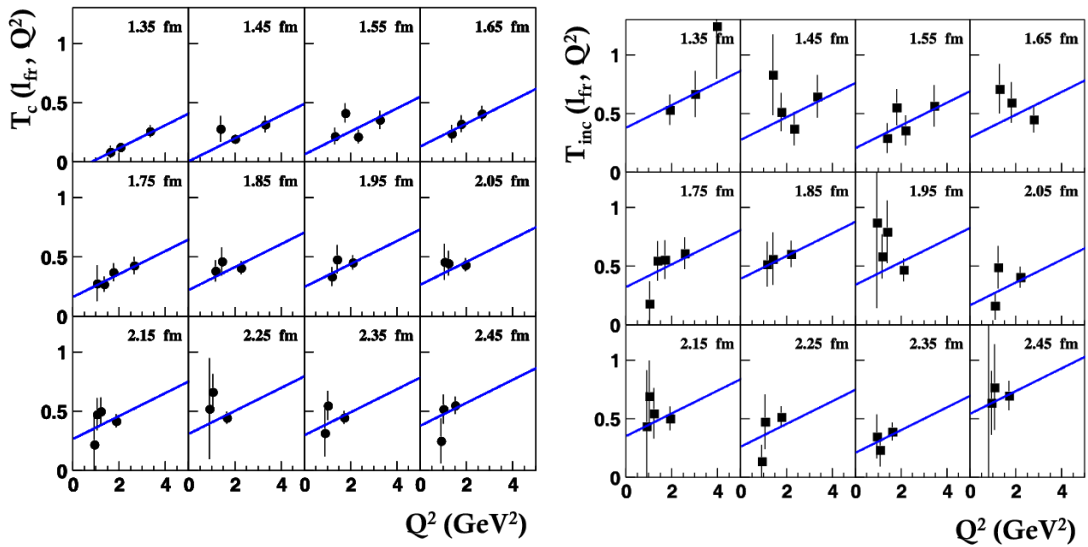

\begin{center}
\begin{minipage}[t]{6 cm}
\centerline{\epsfig{file=hermes-coh,scale=0.45}\epsfig{file=hermes-incoh,scale=0.45}}
\end{minipage}
\begin{minipage}[t]{16.5cm}
\caption{Nuclear transparency as a function of $Q^2$ in specific coherence length bins $l_{fr}$ (as indicated in each panel) for coherent (left) and incoherent (right) $\rho^0$ production on nitrogen \cite{Airapetian:2002eh}. The straight line is the result of the common fit of the $Q^2$-dependence. The error bars include only statistical uncertainties. \label{fig:hermes-trans}}
\end{minipage}
\end{center}
\end{figure}

\subsubsection{Recent $\rho^0$ Meson Electroproduction Experiment}
\label{sec:rho}

\begin{figure}[htb]
\begin{center}
\begin{minipage}[t]{7 cm}
\centerline{\epsfig{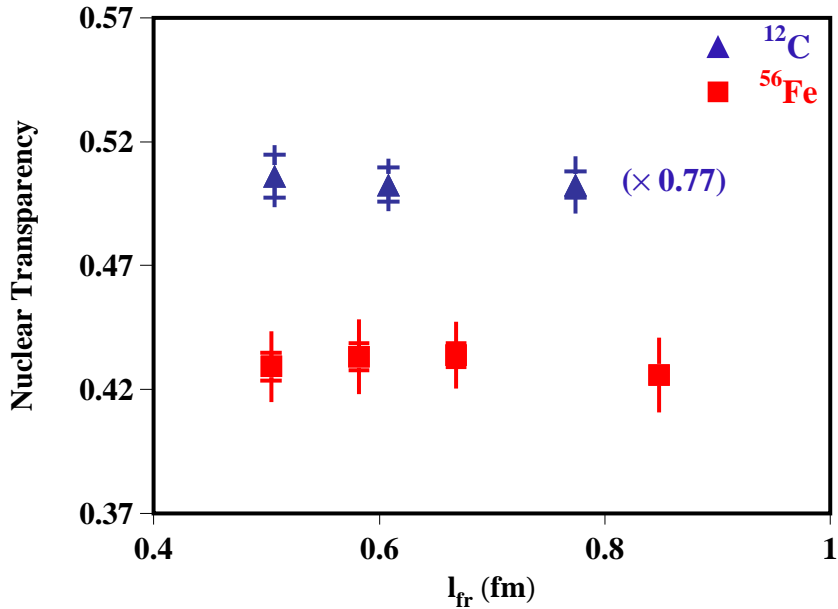}}
\end{minipage}
\begin{minipage}[t]{16.5 cm}
\vspace{-2.5cm}
\caption
    {Nuclear transparency as a function of $l_{fr}$. The inner error bars are the statistical uncertainties and the outer ones are the statistical and point-to-point ($l_{fr}$ dependent) systematic uncertainties added in quadrature.  The carbon data has been scaled by a factor 0.77 to fit in the same figure with the iron data \cite{ElFassi:2012nr}.}  
    \label{fig:clas-lc}
\end{minipage}
\end{center}
\end{figure}

\begin{figure}[htb]
\begin{center}
\begin{minipage}[t]{8 cm}
\centerline{\epsfig{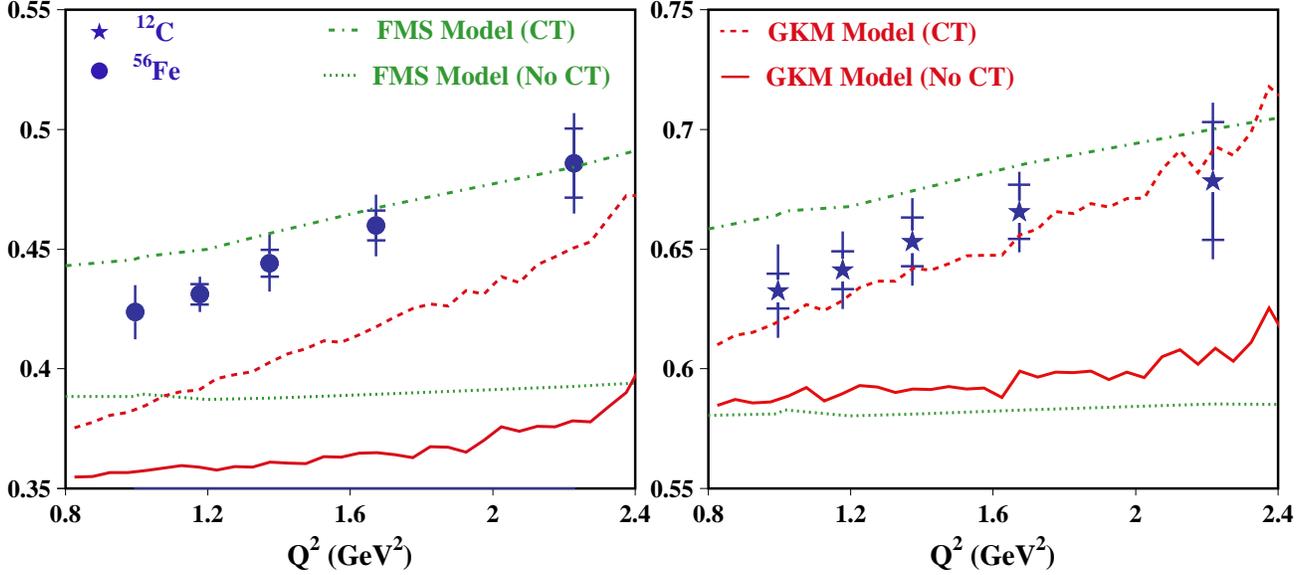}}
\end{minipage}
\begin{minipage}[t]{16.5 cm}
\caption
    {Nuclear transparency as a function of $Q^2$. The inner error bars are statistic uncertainties and the outer ones are statistic and point-to-point ($Q^2$ dependent) systematic uncertainties added in quadrature \cite{ElFassi:2012nr}. The curves are predictions of the FMS \cite{Frankfurt:2008pz} (red) and GKM \cite{Gallmeister:2010wn} (green) models with (dashed-dotted and dashed curves, respectively) and without (dotted and solid curves, respectively) CT. Both models include the pion absorption effect when the $\rho^0$ meson decays inside the nucleus. }  
    \label{fig:clas-q2}
\end{minipage}
\end{center}
\end{figure}

Recently, CLAS collaboration measured the nuclear transparency for incoherent exclusive $\rho^0$ electroproduction off carbon and iron relative to deuterium \cite{ElFassi:2012nr} using a $5~\mbox{GeV}$ electron beam. Both the deuterium target and the solid target (carbon, iron) were exposed to the beam simultaneously to reduce systematic uncertainties in the nuclear ratio and allow high precision measurements. The $\rho^0$ mesons were identified through the reconstructed invariant mass of the two detected pions with 0.6 $< M_{\pi^+ \pi^-} <  1~\mbox{GeV}$. To identify exclusive diffractive and incoherent $\rho^0$ events, a set of kinematic conditions had to be satisfied. The following cuts were applied: $W >  2~\mbox{GeV}$ to suppress pions from decay of resonances,  $-t < 0.4~\mbox{GeV}^2$ to select diffractive events, $-t > 0.1~\mbox{GeV}^2$ to exclude coherent production off the nucleus and $z_\rho = E_\rho/\nu >$ 0.9, where $E_\rho$ is the energy of the $\rho^0$, to select elastically produced $\rho^0$ mesons. The $t$ distributions for exclusive events were fit with an exponential form $Ae^{-bt}$. The slope parameters $b$ for $^2$H (3.59 $\pm$ 0.5), C (3.67 $\pm$ 0.8) and Fe (3.72 $\pm$ 0.6) were reasonably consistent with CLAS \cite{Morrow:2008ek} hydrogen measurements of 2.63 $\pm$ 0.44 taken with $5.75~\mbox{GeV}$ beam energy. The transparencies for C and Fe are shown as a function of $l_{fr}$ in Fig.~\ref{fig:clas-lc}. As expected, they do not exhibit any $l_{fr}$ dependence because $l_{fr}$ is much shorter than the C and Fe nuclear radii of 2.7 and 4.6 fm respectively. Consequently, the photon coherence length effect cannot mimic the CT signal in this experiment. Fig.~\ref{fig:clas-q2} shows the increase of the transparency with $Q^2$ for both C and Fe, indicating the onset of CT phenomenon~\footnote{The calculations are different from those published in~\cite{ElFassi:2012nr} because the authors have corrected a bug in their code for Carbon}. Note that in the absence of CT effects, hadronic Glauber calculations would predict 
practically
no 
$Q^2$ dependence of the nuclear transparency $T_A$ since any $Q^2$ dependence in the $\rho^0$ production cross section would cancel in the ratio and the $\rho-N $ cross section is practically constant in the discussed energy range. The rise in transparency with $Q^2$ corresponds to an $(11 \pm 2.3)\%$ and $(12.5 \pm 4.1)\%$ decrease in the absorption of the $\rho^0$ in Fe and C respectively. The $Q^2$ dependence of the transparency was fitted by a linear form $T_A = a~Q^2 + b$. 

The extracted slopes ``$a$'' for C and Fe are compared to the model predictions in Table~\ref{tab:slopes}. The results for Fe are in good agreement with both Kopeliovich-Nemchik-Schmidt (KNS) \cite{Kopeliovich:2007mm} and Gallmeister-Kaskulov-Mosel (GKM) \cite{Gallmeister:2010wn} predictions, but somewhat larger than the Frankfurt-Miller-Strikman (FMS) \cite{Frankfurt:2008pz} calculations. All models yield an approximately linear $Q^2$ dependence as shown in Fig.~\ref{fig:clas-q2}. The measured slope for carbon corresponds to a drop in the absorption of the $\rho^0$ from $37$\% at $Q^2 = 1~\mbox{GeV}^2$ to 32\% at $Q^2 = 2.2 ~\mbox{GeV}^2$, in reasonable agreement with the calculations. The measured slopes both in CLAS and HERMES are fairly well described  by the KNS model discussed in the section \ref{sec:early_rho}. Within the statistical precision the FMS model is quite successful in reproducing both the slopes and the magnitude of the nuclear transparencies, while taking into account both CT effect and the $\rho^0$ decaying inside the nucleus and the subsequent pion absorption effect. The same model is successful in reproducing the JLab pion electroproduction data discussed in section \ref{sec:pion}. Last, the GKM model is based on BUU transport formalism discussed in section \ref{sec:pion}. The model includes CT effects for $\rho^0$ produced in deep inelastic scattering and seems to produce quite well the carbon data, while it completely fails reproducing iron data. The onset of CT in $\rho^0$ electroproduction seems to occur at lower $Q^2$ than in the pion measurements. This early onset suggests that diffractive meson production might be the optimal way to create small size $q\bar{q}$ pair. The $Q^2$ dependence of the transparency ratio is mainly sensitive to the reduced interaction of the $q\bar{q}$ pair as it evolves into a full-sized hadron, and thus depends strongly on the $\rho$ meson formation time ($l_{c}$) during which the small size configuration's color fields expand to form a $\rho^0$ meson. The formation time used by the FMS and GKM models is 
given by Eq.\ref{lcoh} and changes  between 1.1 and 2.4 fm for $\rho^0$ mesons produced with momenta from $2$ to $4.3~\mbox{GeV}$ while the KNS model uses an expansion length roughly a factor of two smaller. The agreement between the observed $Q^2$ dependence and these models suggests that these assumed expansion distances are reasonable. Having established these features, detailed studies of the theoretical models will allow the first quantitative evaluation of the structure and evolution properties of the small size configurations. Such studies will be further enhanced by future measurements \cite{hafidi06}, which will include additional nuclei and extend to higher $Q^2$ values.

\begin{table}[htb]
\begin{center}
 \caption{\label{tab:table1}Fitted slope parameters of the $Q^2$-dependence of the nuclear transparency for carbon and iron nuclei.~The results are compared with theoretical predictions of KNS~\cite{Kopeliovich:2007mm}, GKM~\cite{Gallmeister:2010wn} and FMS~\cite{Frankfurt:2008pz}.} 
\vspace {0.5cm}
\begin{tabular}{ccccc}
\hline
&\multicolumn{1}{c}{Measured slopes}&\multicolumn{3}{c}{Model Predictions}\\
Nucleus& $\mbox{GeV}^{-2}$&KNS&GKM&FMS\\ \hline
C&$0.044 \pm 0.015_{stat} \pm 0.019_{syst} $& 0.06&0.06 &0.029 \\
Fe&$0.053 \pm 0.008_{stat} \pm 0.013_{syst}$ &0.047&0.047&0.032 \\
 \end{tabular}
\label{tab:slopes}
\end{center}
\end{table} 

\section{Directions for the future studies at JLab}
\begin{figure}[htb]
\begin{center}
\begin{minipage}[t]{7.5 cm}
\centerline{\epsfig{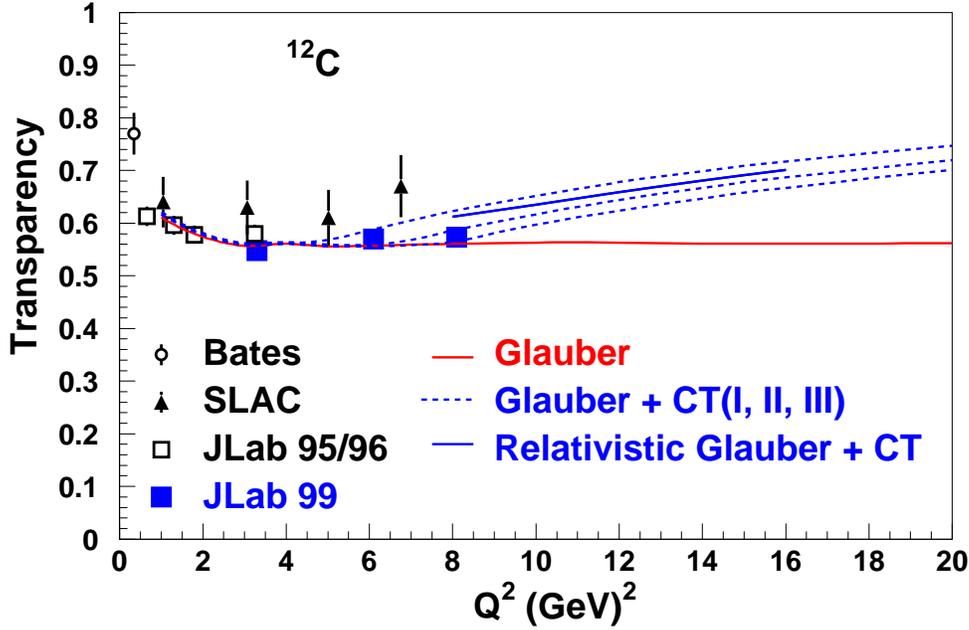}}
\end{minipage}
\begin{minipage}[t]{16.5 cm}
\caption
    {Projected results for the extension of the $A(e,e'p)$ experiment at the upgraded JLab. Nuclear transparency as a function of $Q^2$, for $^{12}$C are shown. The previous data are the same as in Fig.~\ref{fig:eep1}. The error bars represent the quadrature sum of the statistical
and the a 5\% systematic uncertainty. The solid line (red) is the prediction of the Glauber approximation~\cite{vijay:92}. Also shown are the predictions of several CT calculations, dashed blue lines are for CT added to a Glauber calculation~\cite{misak} for three different set of parameters and solid blue is for CT added to a relativistic Glauber calculation.~\cite{Cosyn:2006}.}\label{fig:eep12proj}
\end{minipage}
\end{center}
\end{figure}

\begin{figure}[htb]
\begin{center}
\begin{minipage}[t]{7 cm}
\centerline{\epsfig{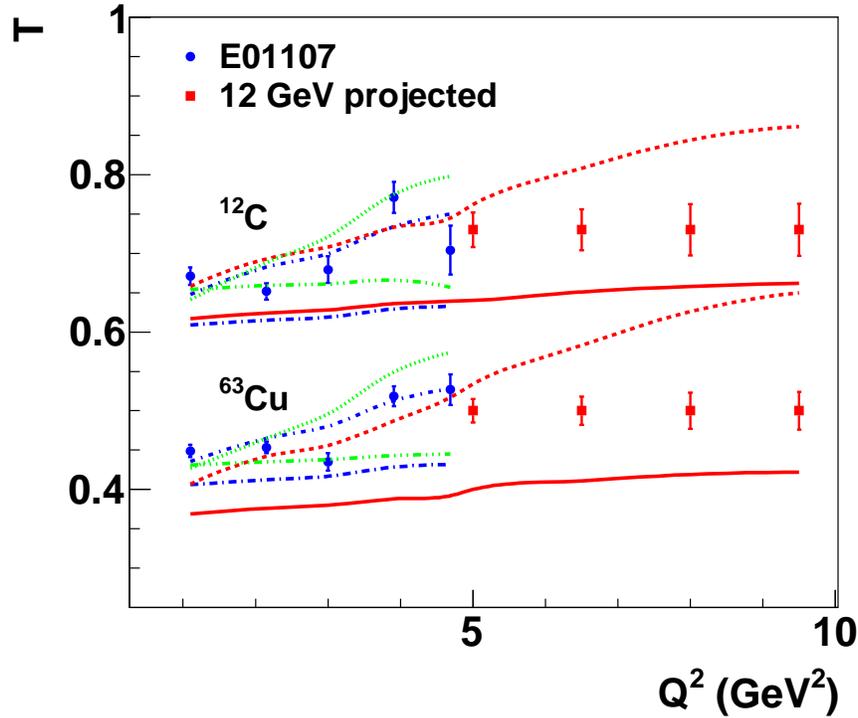}}
\end{minipage}
\begin{minipage}[t]{16.5 cm}
\caption
    {Projected results for the extension of the $A(e,e'\pi)$ experiment at the upgraded JLab.The projected results are shown along with results from previous pion experiment~\cite{Clasie:2007,Qian:2010}. The error bars represent only the statistical uncertainty. All the available calculations are shown along with the projections of Cosyn {\it et. al}~\cite{cosyn2} }  
    \label{fig:eepi12proj}
\end{minipage}
\end{center}
\end{figure}

\begin{figure}[htb]
\begin{center}
\begin{minipage}[t]{7 cm}
\centerline{\epsfig{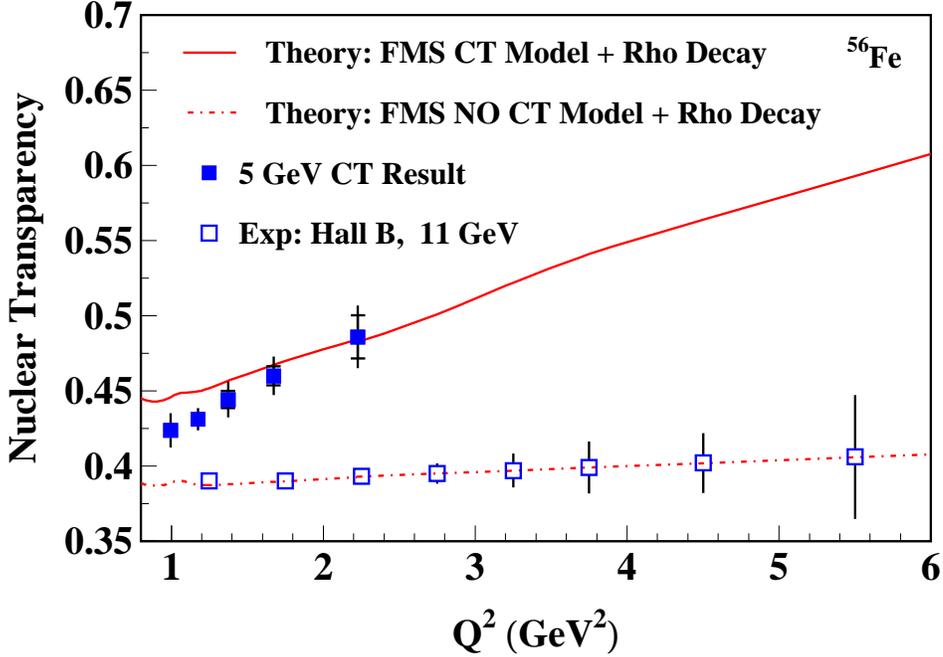}}
\end{minipage}
\begin{minipage}[t]{16.5 cm}
\caption
    {The projected results for the extension of the $A(e,e' \rho ^0)$ experiment at the upgraded JLab. The expected statistical uncertainties are shown for nuclear transparency of iron for the coherence length between 0.4 and 0.5 fm \cite{hafidi06}.}  
    \label{fig:rho-12}
\end{minipage}
\end{center}
\end{figure}

There are already approved  plans for extending CT studies of the $A(e,e'p)$,
 $A(e,e'\pi)$ and $A(e,e'\rho^0)$ reactions to much higher energies following the upgrade of JLab to $12~\mbox{GeV}$. This will finally allow to reach kinematics where $l_{fr}$ is larger than the interaction length for a nucleon/pion in the nuclear media. The extension of the $A(e,e'p)$ experiment will double the $Q^2$ range covered from the current $ Q^2 = 8.0~\mbox{GeV}^2$ to $ Q^2 = 16.0~\mbox{GeV}^2$. At these higher $Q^2$ values CT predictions diverge appreciably from the predictions of conventional 
calculations (see Fig.~\ref{fig:eep12proj}). As mentioned earlier the BNL $A(p,2p)$ data seem to establish a definite increase in nuclear transparency for 
nucleon momenta between about 6 and 10 $~\mbox{GeV}$. For $A(e,e'p)$ 
measurements comparable momenta of the ejected nucleon correspond to 
about $10 < Q^2 < 17~\mbox{GeV}^2$, exactly the range of the proposed extension. Hence, this would unambiguously answer the question
whether one has entered the CT region for nucleons, and help establish
the threshold for the onset of CT phenomena in three-quark hadrons. Moreover,
observation  of CT or lack of CT would help pick out the right explanation for
the energy dependence observed in nuclear transparency from $A(p,2p)$ experiments at BNL. Fig~\ref{fig:eep12proj} shows the projected results for the extension of the $A(e,e'p)$ experiment at the upgraded JLab.
The extension of the $A(e,e'\pi)$ experiment will also double the $Q^2$ range covered from the current $ Q^2 = 5.0~\mbox{GeV}^2$ to $ Q^2 = 10.0 ~\mbox{GeV}^2$. A $Q^2$ dependence of the pion transparency in nuclei may also be introduced by conventional nuclear physics effects at the lower $Q^2$s. Thus one must simultaneously examine both the $Q^2$ and the $A$ dependence of the meson transparency. Several independent calculations~\cite{ralston, miller} predict the CT effect to be largest around $Q^2$ of $10 ~\mbox{GeV}^2$, which is in agreement with the observation of 
nearly
full CT in the Fermilab experiment mentioned above. Using the data collected at $6~\mbox{GeV}$ as a baseline, the new data could help confirm  
 the CT phenomena in mesons and put it on a firm footing. The projected results for the extension of the $A(e,e'\pi)$ experiment at the upgraded JLab is shown in Fig~\ref{fig:eepi12proj}.
%
%
The JLab $12~\mbox{GeV}$ $A(e,e'\rho^0)$ experiment \cite{hafidi06} will extend the maximum $Q^2$ reach from $2.2$ to $5.5~\mbox{GeV}^2$, which will allow for significant increase in the momentum and energy transfer involved in the reaction. Therefore, one expects to produce smaller configurations that live longer: the optimum parameters for CT studies. Several nuclei including deuterium, carbon, iron and tin will be studied. Measurements with different nuclei sizes are important for quantitative understanding of the small size configuration's formation time and its interaction in the nuclear medium. The dependence of the nuclear transparency on the coherence length will be measured for $l_{fr}$ range up to 2.5 fm. The measurements will be performed for fixed coherence length. Fig. \ref{fig:rho-12} shows the projected statistical uncertainties for iron with predictions from the FMS model \cite{Frankfurt:2008pz}. 
%
%

A complementary strategy is to use processes where multiple rescatterings dominate in light  nuclei ($^2$H,$^3$He) which allows to suppress the expansion effects. An additional advantage of these processes is that one can use for the calculations generalized eikonal approximation (GEA), see review in \cite{Sargsian:2001ax}. The recent measurements of the deuteron break up 
\cite{Boeglin:2011mt} at $Q^2 =3.5~\mbox{GeV}^2$  show a good agreement with the GEA prediction including the rescattering kinematics which is most sensitive to the CT effects.
In particular,  these reactions are well suited to search for a precursor  of CT - suppression of the configurations in nucleons with  pion cloud in the hard processes like the nucleon form factors at relatively small $Q^2\ge 1 ~\mbox{GeV}^2$ - chiral transparency 
\cite{Frankfurt:1996ai}. The simplest reaction of this kind is production of a slow $\Delta$ isobar in the process $\mbox{e}^2 + \mbox{H} \to \mbox{e+p+}\Delta^0$ which should be suppressed in the chiral transparency regime.
Another option to test chiral dynamics would be the study of quasielastic production of $N\pi$ system with mass close to the threshold.  
The use of the large $Q^2$ soft pion theorem suggests  \cite{Pobylitsa:2001cz} that at large $Q^2$ the process is dominated by the transition of $3 q$ system into a $N\pi$ system. Hence the strength of absorption in this case for the scattering off nuclei with $A\le 12$ when the coherence length is comparable to $R_A$  should be the same as in the case of nucleon production, while naively the absorption rate should be much larger as it would correspond to the sum of the $\pi N $ and $NN$  cross sections.

Two other examples are  (i) large angle $\gamma + N \to "meson" +N$ reaction in nuclei where one should first look for  
a change of A-dependence from $\propto A^{1/3}$ to $\propto A^{2/3}$ already in the region where expansion effects are large due to transition from the vector dominance regime to the regime of point-like photon interaction in which the photon penetrates to any point in the nucleus, (ii)  A-dependence of virtual Compton scattering, namely at what $Q^2$  transition from  vector dominance regime to the CT regime occurs. HERMES data~\cite{Airapetian:2009bi} are consistent with  predictions of \cite{Guzey:2003jh} for the incoherent channel based on the CT and the closure approximation and for the A-dependence of the coherent cross section for $A\ge 4$, but not for the $^4He/p$ ratio. Note however that  the subtraction of incoherent contribution  differs significantly for proton and heavier targets. 
 
 \section{Directions for the study of color transparency with hadron beams}
\label{sec:hadro-beam}	
	
In addition to the electroproduction and photoproduction programs at JLab,  it is very important to have a parallel program of studies with hadron beams. Such a program in principle would be possible both at the Japan Proton Accelerator Research Complex (J-PARC)~\cite{jparc} and at the Facility for Anti-proton and Ion Research (FAIR)~\cite{fair}.
 In the case of FAIR, the configuration of the PANDA experiment appears to be especially promising. 
 
\subsection{Study of the two body, large angle processes with nucleon targets} 
  
Understanding the large angle exclusive processes remains one of the challenges for pQCD.  The PANDA experiment at  FAIR, which will use  antiproton beams on internal targets,  will have excellent acceptance for numerous large angle processes, ranging from the simplest processes of $p\bar p \to p\bar p, \pi \pi, K \bar K$ to the processes of production of multi-particle  states such as baryon - antibaryon  states $(\Sigma^+ \bar{\Sigma}^-,...) $ and  meson pairs $(\pi^-\rho^+,...)$. As a result it  will be able to study many di-meson and baryon - antibaryon states over a much larger range of energies and channels than covered by previous measurements~\cite{White:1994tj} ($p_{\bar p} $ = 6 \& 9.9 $\mbox{GeV}$). Such experiments will allow verification of the observations of Ref.~\cite{White:1994tj}, that cross sections for the processes where quark exchanges are allowed are much larger than those where such exchanges are not allowed. It will also provide confirmation of whether the   ratio of $\bar p p $ and $pp$ elastic scattering continues to fall with increasing momenta for momenta $>  6~\mbox{GeV}$ or if it becomes momentum independent. A flat $\bar p p $ and $pp$ ratio would indicate that the diagrams with gluon exchanges also satisfy dimensional counting rules.
  
 Another puzzle to address  is the oscillation of the differential cross section of the elastic $pp$ scattering at large angles  around a smooth quark counting inspired parametrization. Are these oscillations present  in $pp$ scattering away from $\theta_{c.m.}=90^\circ$? Are they present in any of the $p \bar p$ channels? Other channels, such as those involving production of 
  $\Delta$-isobars, non-resonant $\pi N$ production, etc are practically unknown. Another gap in the knowledge is $pn$ scattering which could be studied using 
the $^2$H targets. There is a  suggestion that the measurement of the $pn/pp$ ratio may provide an insight on the SU(6) structure of the nucleon wave function 
at large x \cite{Sargsian:2009}. Overall, comparing all these channels  in $pN$ and $\bar pN$ scattering may lead to a breakthrough in the understanding of hard  two body reactions.

\subsection{Basic reactions with nuclear targets}

The use of nuclear targets allows experimental confirmation whether basic two body $a+ N  \to c+d$  reactions are dominated by small interquark distances.
Presence of CT in the pion electroproduction and the change in the transparency of $pA\to pp (A-1)$ process for $p_N \ge 8~\mbox{GeV}$ \cite{Leksanov:01} indicate that  experiments with antiprotons of energies above $6 - 8~\mbox{GeV}$ where a pair of pions is produced in the final state will be able to verify whether the annihilation process is dominated by the contribution from small size configurations, by measuring the transparency ratio; $T=\sigma(\bar p A \to \pi^+\pi^- (A-1))/\sigma(\bar p p \to \pi^+\pi^- )$. An increase of $T$ with energy will signal onset of the CT regime. The JLab results on the pion dynamics (both current and from the upgraded JLab which will be available about the same time as the completion of FAIR) will allow an unambiguous interpretation of the data.  Since the elementary cross section drops very rapidly with increase of energy for 
a fixed value of $\theta_{c.m.}$ one would have to switch from the study of transparency at fixed  $\theta_{c.m.}\sim 90^o$ to study of transparency at large but fixed $t$. In this case one expects an increase of the  transparency with 
increase of the incident energy due to a gradual  freezing of the wave function of the projectile and the faster of two final particles, for example in Ref.~\cite{Strikman:1998sf,Miller:2010eh} where both $p,2p$ and $\pi, \pi N$ were considered at high energy kinematics.

\subsection{Rescattering kinematics}
 
Another method to study CT phenomenon in detail is the use of the rescattering patterns in exclusive reactions. For example, $p(\bar p)~+~^2H~\to~p(\bar p)~p  + n $, where the neutron is slow \cite{Frankfurt:1996uz}. This reaction can be separated from background processes without a need to detect the neutron, provided a $\sim$ 1\% momentum resolution for the fast particles is achieved and a veto for the pion production is implemented. Advantage of this reaction is that one can study the patterns of the rescattering in detail, with characteristic distances between the centers of about $1$~fm, where expansion effects are strongly suppressed.  Similarly one can use scattering from a $^3He$ target in the rescattering 
kinematics -- $h +~^3He \to h  + p + ^2H$. 
 
\subsection{Testing chiral dynamics effects}

As we discussed in Section 4, studies~\cite{Pobylitsa:2001cz} suggest that  one can use soft pion theorems in the hard processes to explore the dynamics of the baryon production by virtual photons in the process $\gamma^*N\to N\pi$ with
 $M_{\pi N}-M_{\pi} -M_N\le 100~\mbox{MeV}$, leading to the expectation of 
larger nuclear transparency than in a  naive model where $N$ and $\pi$ 
propagate incoherently. Similar approach is possible in the case of hadronic 
processes. One can explore large angle reactions  like  
\begin{equation}         
\bar p  (p) A \to  N\pi ( \bar p\pi ) + p + (A-1),
 \end{equation}
 in the kinematics where invariant mass of $N\pi$ is close to the threshold. 
 If  the process proceeds through the three quark stage with subsequent 
transition $3q \to N\pi$,  nuclear transparency will be the same as in the 
process without pion emission, though if there is no chiral stage, the  
absorption will be much larger. Another method to probe the presence of the  
pion cloud close to the interaction point is to look for the charge exchange 
processes like production of slow $\Delta$ in the 
process $p +~^2H \to pp + \Delta^0$ \cite{Frankfurt:1996ai}.
  
 \subsection{Probing properties of hadron containing charm quark in $\bar p A$ collisions}

Charmonium states can  be produced in the nuclear media in the same resonance reactions as the ones  which are studied with a $^2$H target: $\bar p p \to J/\psi,\chi_c,...$. A possibility to use these processes for studies of CT was first suggested in Ref.~\cite{Brodsky:1988xz}. A subsequent detailed analysis of the effects of CT and Fermi motion was performed in Ref.~\cite{Farrar:1989vr}. It 
was found that CT effects due to squeezing of the incoming antiproton are 
small since the incident energy is small, leading to a short coherence length 
for $\bar p$ and the produced charmonium.   The effect of the Fermi  motion 
is large but calculable with sufficient precision. What is unique about these 
reaction is that at these energies charmonium is formed very close to the 
interaction point  $\le $ 1 fm. Hence they allow verification of the main 
premise of CT, that small objects interact with nucleons with small 
cross sections. The ability to select states of varying 
size: $J/\psi, \chi_c, \psi'$ will allow studies of how the strength of the 
interaction depends on the transverse size of the system up to the sizes 
comparable to the pion size.  The filtering of $\bar c c $ configurations of 
smaller transverse size leads  to the   $A$-dependent polarization of 
the $\chi$ states \cite{Gerland:1998bz}. Study of this phenomenon is also 
important for understanding the dynamics of charmonium production in 
heavy ion collisions.

There are several other channels sensitive to the dynamics of charm interaction. Near the peak of  resonance production of $J/\psi$ one can look for 
production of $D, \bar D, \Lambda_c, \psi'$  resulting from the interactions 
of the $J/\psi$ with the nucleons of the target nucleus \cite{Gerland:2005ca}. 
Similar measurements are possible near the $\psi'$ resonance. One can use 
excitation of the charmonium resonances above the $\bar DD $ threshold 
and look for softening of the $D$-meson spectrum with $A$ which is a measure of 
the $D-N$ interaction. Such a measurement would be hardly possible in any 
other processes.

\subsection{Opportunities with COMPASS}
	
We have argued above that in the investigation of the $s, t$ range, in which the interaction of small size configuration dominates  the cross section of the large 
angle $2\to 2$ processes,  the use of CT is  hampered by the presence of the 
rather fast space-time  evolution of the wave packages of the colliding 
hadrons. Even if the cross section of, for example $\pi N$, scattering is dominated by the contribution of point-like configurations at $s= 9 ~\mbox{GeV}^2$, and $\theta_{c.m.}=90^\circ$, it would be impossible to observe a strong CT effect in the $(\pi, \pi N)$ reaction since the coherence length will be too small.

A solution to  this conundrum  was proposed in Ref.~\cite{Kumano:2009ky}. The idea is to use $2\to 3 $ processes \cite{Kumano:2009he} where a high energy 
hadron "b" collides with the target "a" (nucleon) producing a recoil 
particle "e" and two leading hadrons "c" and "d" with 
\begin{equation}
s_{eff}=(p_c+p_d)^2 \ll s, t_{eff}= -(p_b- p_d)^2 /s_{eff} = const \sim 1/2, t=(p_a-p_e)^2=const
\end{equation}

\begin{figure}[htb]
\begin{center}
\begin{minipage}[t]{4 cm}
\centerline{\hspace{-2.5 cm}\epsfig{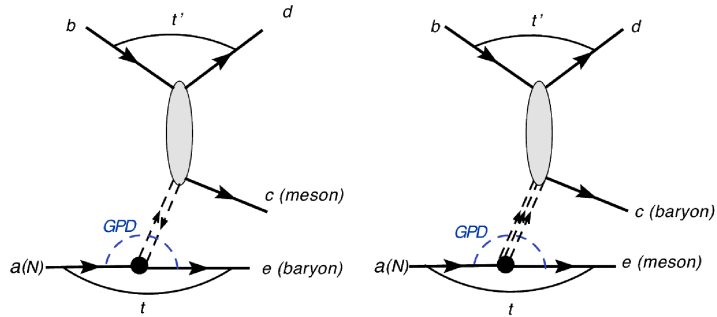}}
\end{minipage}
\begin{minipage}[t]{16.5 cm}
\vspace{-1 cm}
\caption{Diagrams for the $2\to 3$ processes with $q\bar q$ and $3q$ exchange in t-channel. \label{2to3}}
\end{minipage}
\vspace{-0.5 cm}
\end{center}
\end{figure}

If the subprocess producing "c + d" system is hard, the exchange in the $t$-channel is by either $q\bar q $ pair - for a reaction like $\pi^- p \to \pi^-\pi^{+} n, \pi^- p \to \pi^-\pi^- \Delta^{++}$  or by three quarks  ($p+p  \to p+\pi^+ n$), see for example Fig.~\ref{2to3}, and 
the vertex for the emission of  $q\bar q (3q)$  is given by the 
corresponding generalized parton distribution (GPD). One also expects that a 
factorization theorem similar to the case of exclusive meson production in 
DIS is valid~\cite{Kumano:2009he} for these processes.
Since $s_{eff}$ is small we can consider the limit when the longitudinal components of the  momenta of both fast hadrons "c" and "d"  are comparable and $\sim p_a/2$. In this limit $l_{c}$ can be chosen to be $\gg  R_A$ so the projectile and constituents forming  "c" and "d" are frozen during the propagation through the nucleus.

\begin{figure}[htb]
\begin{center}
\begin{minipage}[t]{7 cm}
\vspace{-0.4 cm}
\centerline{\epsfig{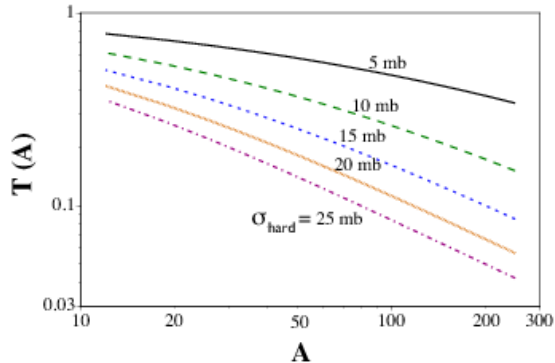}}
\end{minipage}
\begin{minipage}[t]{16.5 cm}
\vspace{-7 cm}
\caption{$A$-dependence of $T(A)$ for different values 
         of $\sigma_{\rm hard}$ for the $\pi A \to \pi \pi A^*$ reaction~\cite{Kumano:2009ky}.}\label{diagram:Fig3}
\end{minipage}
\vspace{-6 cm}
\end{center}
\end{figure}

\begin{figure}[htb]
\begin{center}
\begin{minipage}[t]{7 cm}
\centerline{\epsfig{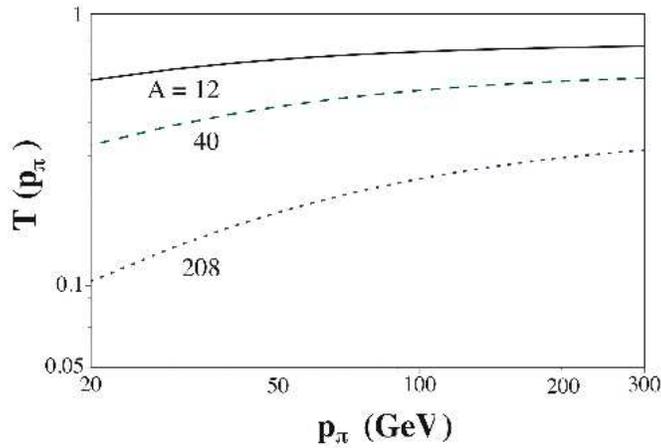}}
\end{minipage}
\begin{minipage}[t]{16.5 cm}
\caption{The pion momentum ($p_{\pi}$) dependence of the nuclear transparency~\cite{Kumano:2009ky}.  }\label{diagram:Fig4}
\end{minipage}
\end{center}
\end{figure}

Let us consider for simplicity the process $\pi^- A \to \pi^-\pi^- A^*$ for 
$p_{\pi} \sim 200~\mbox{GeV}$ (data on production of the leading hadrons in $\pi^-A $ scattering were collected by COMPASS which so far focused on the study of the diffractive channel). Note that due to the color transparency effect, it 
is sufficient for the excitation energy of the recoil system to be below $\sim 1 ~\mbox{GeV}$, which is quite feasible with the available energy resolution of 
COMPASS. One finds that the transparency in the discussed kinematics has indeed a  minor  sensitivity to the expansion effects and high sensitivity to the strength of the interaction of the $q\bar q$ system in the interaction point, $\sigma_{hard}$. One can see from Fig.~\ref{diagram:Fig3} that even reduction of the strength of the 
interaction by a factor of two would very strongly change the 
nuclear transparency. It would be interesting to investigate the transparency as a function of the transverse momentum of pions (remember that in the case of pion dissociation into two jets CT was observed already at $p_t\sim  1.5~\mbox{GeV}$). If 
the CT effect is observed one would be able to investigate the space time evolution of the high energy wave packages by changing the initial pion momentum and 
keeping $s_{eff}$ and $t_{eff}$ the same, for example, see Fig.~\ref{diagram:Fig4}~\cite{Kumano:2009ky}. Also it would open a new way for measuring 
non-vacuum exchange  GPDs of nucleons and will allow for the first time the measurement of GPDs 
of other $t$-channel hadrons. It will also allow investigation of the large angle meson - meson scattering.
%
%

\subsection{Opportunities with heavy ion collisions at the LHC}

Heavy ion collisions at the LHC provide a unique opportunity to study CT in the coherent and quasielastic photoproduction of $J/\psi$ and $\Upsilon$ via selection of the so called ultraperipheral processes in which one of the nuclei serves as a source of quasireal  photons, for the review see \cite{Baltz:2008}. This would allow to extent substantially the range of the invariant energies  as compared 
to the ones studied at FNAL. At these energies one expects to observe the  transition from the regime of the nearly complete nuclear transparency to the regime of a weaker nuclear dependence. In such a regime a transition is expected both in the leading twist approximation (the leading twist nuclear shadowing) and in the models where interaction of small dipoles  starts to approach black disk regime at small impact parameters at very high energies.
Further studies along these lines would be possible at an electron - ion collider as well.

\section{Conclusions}
To summarize, the high energy CT is well established and will be further studied at various future facilities.  There is evidence for the onset of CT regime in exclusive meson production at JLab at $Q^2$ of few $\mbox{GeV}^2$. It is likely that JLab experiments  at $12~\mbox{GeV}$ will  observe significant CT effects  for the processes with meson production and will provide a  decisive test of whether nucleon form factors at $Q^2\sim 15 ~\mbox{GeV}^2$ are dominated by PLC or mean field configurations.  CT will also help establish the interplay between soft and hard physics for many other exclusive large momentum transfer processes at JLab, EIC, LHC  as well as at  hadronic factories J-PARC, FAIR and in the fixed target experiments at CERN (COMPASS).

\section{Acknowledgments}
This work was supported by the U.S. Department of Energy, Office of Nuclear Physics, under contracts No.~DE-AC02-06CH11357 and ~DE-FG02-07ER41528.
We thank R. Ent, G. Miller and M. Sargsian for numerous discussions. MS thanks L. Frankfurt and M. Zhalov for numerous discussions. We also thank M. Zhalov for help with generating Figs. 10 and 12, X. Qian for help with Figs. 14 and 15 and L. El Fassi for help with Figs. 21 and 24.


\end{document}